\documentclass[sigconf]{acmart}

\PassOptionsToPackage{protrusion, expansion}{microtype} % for better typograph

\usepackage{amsmath, amsthm}
    
\usepackage{graphicx, xcolor} % xcolor instead of color
\usepackage{hyperref, url}
\usepackage{adjustbox}
\usepackage{array}
\usepackage{booktabs, multirow}
\usepackage{caption}
\usepackage{subcaption}
\usepackage{enumitem}
\usepackage{lipsum}
\usepackage{setspace}
\usepackage[normalem]{ulem}
\usepackage{xcolor}

\usepackage{tikz}
\newcommand*\Circled[1]{% require `tikz`
	\tikz[baseline=(char.base)]{\node[
        shape=circle, draw=none,  thick, 
        fill=gray!40,inner sep=0.6pt] (char) 
    {\textcolor{black}{\sffamily#1}}; 
}}
\usepackage[ruled, linesnumbered, norelsize]{algorithm2e}

\usepackage{footnote}
\makesavenoteenv{tabular}
\makesavenoteenv{table}
\usepackage[zerostyle=d]{newtxtt}

\setcopyright{rightsretained}

\copyrightyear{2022}
\acmYear{2022}
\setcopyright{rightsretained}
\acmConference[ICS '22]{2022 International Conference on Supercomputing}{June 28--30, 2022}{Virtual Event, USA}
\acmBooktitle{2022 International Conference on Supercomputing (ICS '22), June 28--30, 2022, Virtual Event, USA}
\acmDOI{10.1145/3524059.3532362}
\acmISBN{978-1-4503-9281-5/22/06}

\begin{document}

\fancyhead{}

\title{CEAZ: Accelerating Parallel I/O via Hardware-Algorithm Co-Designed Adaptive Lossy Compression}

\settopmatter{authorsperrow=3}

\newcommand{\AFFIL}[4]{%
     \affiliation{%
         \institution{\small #1}
         \city{#2}\state{#3}\country{#4}
     }
     }

\author{Chengming Zhang}{\AFFIL{Washington State University}{Pullman}{WA}{USA}}
\email{chengming.zhang@wsu.edu}

\author{Sian Jin}{\AFFIL{Washington State University}{Pullman}{WA}{USA}}
\email{sian.jin@wsu.edu}

\author{Tong Geng}{\AFFIL{Pacific Northwest National Laboratory}{Richland}{WA}{USA}}
\email{tong.geng@pnnl.gov}

\author{Jiannan Tian}{\AFFIL{Washington State University}{Pullman}{WA}{USA}}
\email{jiannan.tian@wsu.edu}

\author{Ang Li}{\AFFIL{Pacific Northwest National Laboratory}{Richland}{WA}{USA}}
\email{ang.li@pnnl.gov}

\author{Dingwen Tao}{\AFFIL{Washington State University}{Pullman}{WA}{USA}}
\authornote{Corresponding author: Dingwen Tao, School of Electrical Engineering and Computer Science, Washington State University, Pullman, WA 99163, USA.}
\email{dingwen.tao@wsu.edu}

\begin{abstract}
As HPC systems continue to grow to exascale, the amount of data that needs to be saved or transmitted is exploding. 
To this end, many previous works have studied using error-bounded lossy compressors to reduce the data size and improve the I/O performance. 
However, little work has been done for effectively offloading lossy compression onto FPGA-based SmartNICs to reduce the compression overhead.
In this paper, we propose a hardware-algorithm co-design for an efficient and adaptive lossy compressor for scientific data on FPGAs (called CEAZ), which is the first lossy compressor that can achieve high compression ratios and throughputs simultaneously.
Specifically, we propose an efficient Huffman coding approach that can adaptively update Huffman codewords online based on codewords generated offline, from a variety of representative scientific datasets. 
Moreover, we derive a theoretical analysis to support a precise control of compression ratio under an error-bounded compression mode, enabling accurate offline Huffman codewords generation. This also helps us create a fixed-ratio compression mode for consistent throughput. 
In addition, we develop an efficient compression pipeline by adopting cuSZ's dual-quantization algorithm to our hardware use cases. 
Finally, we evaluate CEAZ on five real-world datasets with both a single FPGA board and 128 nodes (to accelerate parallel I/O).
Experiments show that CEAZ outperforms the second-best FPGA-based lossy compressor by $2.3\times$ of throughput and $3.0\times$ of ratio. It also improves MPI\_File\_write and MPI\_Gather throughputs by up to $28.9\times$ and $37.8\times$, respectively.

\end{abstract}




\maketitle

\setlength{\textfloatsep}{6pt}
\setlength\abovecaptionskip{3pt}
\setlength{\abovedisplayskip}{2pt}
\setlength{\belowdisplayskip}{2pt}
\setlength{\abovedisplayshortskip}{2pt}
\setlength{\belowdisplayshortskip}{2pt}

\section{Introduction}
\label{sec:introduction}

Today's HPC applications can generate large volumes of scientific data for post-hoc analysis and visualization. 
%Moreover, as supercomputers continue to grow to exa-scale, the amount of data that needs to be snapshot or transmitted is exploding.
However, since the development of storage and networking hardware is much slower than that of computing power and memory capacity \cite{cappello2019use}, the I/O and network bandwidth are becoming the main bottlenecks for HPC applications to achieve high performance on a large scale.
I/O and communication costs can quickly overwhelm the overall performance as parallel computers grow towards exascale. 
For instance, a well-known cosmological simulation code Nyx~\cite{almgren2013nyx} can generate up to 2.8 TB of data for a single snapshot under a simulation resolution of $4096^3$, requiring to save a total of 2.8 PB data, when running the simulation for 5 times with 200 snapshots dumped per run.

Such a large amount of data is often generated in a parallel manner from a scaling number of ranks, on which each holds a proportion of the data and must introduce an extra collective communication to dump the entire snapshot to the file system. This process takes an unprecedented challenge to I/O bandwidths and storage systems on today's HPC systems~\cite{liangerror,wan2017comprehensive,wan2017analysis,cappello2019use}.
Therefore, it is urgent to develop effective data reduction methods to reduce the size of data movement between memories and storage systems such as parallel file systems. 

%A straightforward approach is decimation, i.e., storing one snapshot every several timesteps during the simulation. However, even with decimation, there are still a number of timesteps to store, and a large amount of data in one timestep can still overwhelm the storage capacity and I/O bandwidth, not to mention this approach may cause a significant loss of valuable information. 
One of the most effective ways to address this challenge is using data compression. The data partition in each rank is compressed before sending it to the storage system via an interconnected network. This can reduce both I/O overhead and storage consumption. 
%(1) provide less amount of data to be transferred thus less I/O performance overhead, and (2) generate compressed dataset to the storage systems and reduces storage consumption.
However, traditional lossless compression can only provide a limited compression ratio to the scientific dataset by usually up to $2\times$~\cite{son2014data}. Thus, error-bounded lossy compressors such as SZ~\cite{tao2017significantly, di2016fast, liangerror}, ZFP~\cite{zfp}, and MGARD~\cite{ainsworth2017mgard} have been developed to provide a much higher compression ratio while only introducing controllable distortion of data.
Many prior studies have demonstrated the effectiveness of using those error-bounded lossy compressors for scientific data reduction~\cite{di2016fast,tao2017significantly,zfp,liangerror,lu2018understanding,luo2019identifying,tao2018optimizing,cappello2019use,jin2020understanding,grosset2020foresight,jin2021adaptive} and improving I/O performance. \cite{gok2018pastri,wu2019full,poppick2020statistical}.

While error-bounded lossy compressors on CPUs can provide a high compression ratio, their low throughputs unavoidably cause relatively high performance overheads to applications, which often offset the performance benefit from saving and loading compressed data of less sizes.
Thus, we need to develop a high-throughput lossy compressor to effectively accelerate parallel I/O for HPC applications. Recently, all SZ, ZFP, and MGARD teams started to develop and release their GPU and/or FPGA implementations.
On one hand, GPU's massive SIMT parallelism enables high throughput. However, during the lossless compression step of SZ algorithm, Huffman encoding and decoding \cite{Huffman-original} results in a random memory access pattern \cite{tian2020cusz}. This causes serious divergence issues, inevitably leading to low GPU memory bandwidth utilization and performance.
On the other hand, FPGAs offer many advantages, such as configurability, high energy efficiency, low latency, and low price \cite{geng2019o3bnn}, and have been a viable and popular option at scales for smart network interface cards (SmartNICs)~\cite{firestone2018azure, putnam2014reconfigurable}, which are being increasingly used in data centers to offload networking functions from host processors \cite{haghi2020fpgas}. Thus, this makes FPGA-based SmartNICs ideal platforms to offload compression and accelerate I/O.

The state-of-the-art FPGA-based lossy compressor is only capable of about 8 GB/s throughput~\cite{sun2020burstz}, which is still much lower than the throughput of PCIe3/4 and InfiniBand. This precludes their use in real application scenarios.
There are several challenges to implement a high-throughput lossy compression on FPGAs with a relatively high compression ratio: \Circled{1} Most of lossy compression algorithms involve multiple stages which have strong data dependency. \Circled{2} It is infeasible to simply add more compression pipelines on FPGAs with limited resources. 
%: (1) FPGAs usually have much lower clock frequencies (e.g., 450 Mhz) compared to CPUs and GPUs. It is difficult to use such low frequencies to achieve relatively high throughputs. (2) The limited resources on a single FPGA chip prevent us from increasing the throughput by simply adding more compression pipelines. 
Thus, we need to deeply optimize the algorithm to effectively utilize the hardware resources.

To address these challenges, in this work, we focus on designing an efficient lossy compression algorithm that is suitable for FPGA hardware, and offload it onto FPGA-based SmartNICs to accelerate parallel I/O. 
Specifically, we propose a hardware-algorithm \underline{c}o-designed \underline{e}fficient and \underline{a}daptive lossy compressor (\underline{z}ip) (CEAZ)\footnote{The code of CEAZ is available at \url{https://github.com/szcompressor/CEAZ}.}, which is the first lossy compressor to achieve high compression ratio and throughput simultaneously. 
CEAZ adopts a dual quantization strategy~\cite{tian2020cusz} to completely remove data dependency and instantiates multiple pipelines to process input data in parallel. Unlike cuSZ that implements dual-quantization using massive GPU threads, we implement the dual quantization in CEAZ using a pipelined manner, which is more suitable for FPGA architecture. Moreover, different from existing FPGA-based lossy compressors such as GhostSZ \cite{xiong2019ghostsz} and waveSZ \cite{tian2020wavesz} that statically build trees in Huffman coding for every data chunk, CEAZ dynamically determines whether to update codewords by building a new tree or use previous/offline codewords according to the distribution of current symbol frequencies, leading to a high throughput.
%Different from existing works, CEAZ has the following three main innovations. (1) We adopt a dual quantization strategy~\cite{tian2020cusz} to completely remove data dependency and instantiates multiple pipelines to process input data in parallel. Unlike cuSZ that implements dual-quantization using massive GPU threads, we implement dual quantization in a pipelined manner in CEAZ, which is more suitable for FPGAs architecture. (2) We adaptively build a new Huffman tree to update codewords only when the distribution of symbols' frequencies meets a specific criterion. However, GhostSZ and waveSZ will build Huffman trees no matter what the situation. (3) We will directly compress symbols generated by previous steps, which ensures throughput.
Thus, CEAZ can efficiently and effectively reduce the data size and significantly increase the parallel I/O performance.
In addition, to the best of our knowledge, CEAZ is also the first lossy compressor to enable fixed-rate compression for prediction-based compression workflow.
Our contributions are summarized as follows:
\begin{itemize}[noitemsep, topsep=2pt, leftmargin=1.3em]
    \item We propose an efficient Huffman coding approach that adaptively updates Huffman codewords online based on our offline Huffman codewords, which are generated from a variety of representative scientific datasets. It can reduce the data dependency in Huffman coding and dramatically improve the compression throughput on datasets of various sizes.
    \item We derive a theoretical analysis to support a precise control of compression ratio under the error-bounded compression mode, which can align quantization-code histograms of different datasets and enable an accurate generation of offline Huffman codewords. It also helps us develop a fixed-ratio compression mode, which is important to guarantee a consistent throughput in data transfer. Our work is the first prediction-based lossy compressor to enable fixed-rate compression.
    \item We develop an efficient compression pipeline by adapting the dual-quantization algorithm to our hardware use case.
    \item We evaluate CEAZ with five real-world scientific datasets in both serial and parallel processing. Experiments show that CEAZ outperforms the state-of-the-art solution by $2.3\times$ in throughput and $3\times$ in ratio on a single FPGA board.
    Moreover, CEAZ can improve the MPI\_File\_write and MPI\_Gather throughputs by up to $28.92\times$ and $37.8\times$, respectively, with 128 nodes from the Summit supercomputer. 
\end{itemize}

The rest of this paper is organized as follows. In Section 2, we discuss the background and research challenges. In Section 3, we present the design of CEAZ. In Section 4, we evaluate CEAZ on six scientific datasets and present our results. In Section 5, we summarize our work and discuss future work. 

\section{Background and Motivation}
\label{sec:background}
In this section, we first present background information about scientific data compression, FPGA-based lossy compression, and MPI communication and I/O. We then discuss the challenges and motivations for our research.

\subsection{Floating-Point Data Compression}
Floating-point data compression has been studied for decades. The data compressors can be split into two categories: lossless compression and lossy compression. In comparison to lossy compression, lossless compression such as FPZIP~\cite{lindstrom2006fast} and FPC~\cite{FPC} can only provide limited compression ratios (typically up to 2:1 for most scientific data) due to the significant randomness of the ending mantissa bits, especially for large scientific floating-point data~\cite{son2014data}.

Lossy compression, on the other hand, can compress data with little information loss in the reconstructed data. Compared to lossless compression, lossy compression can provide a much higher compression ratio while still maintaining useful information for scientific discoveries. 
Many lossy compressors supporting floating-point data were proposed and designed for visualization.
Thus, many lossy compressors employ techniques directly inherited from lossy compression of images, such as variations of wavelet transforms, coefficient prioritization, and vector quantization. 
%Lossy compressors for image processing are designed and optimized considering human perception, such as JPEG2000~\cite{taubman2012jpeg2000}.
While such compressors may be adequate for visualization, they do not provide error controls on demand for scientific studies. 

In recent years, a new generation of lossy compressors for scientific floating-point data has been proposed and developed, such as SZ~\cite{di2016fast, tao2017significantly, liangerror,tian2020cusz} and ZFP~\cite{zfp,cuZFP}.
Both lossy compressors can provide multiple compression modes, such as error-bounded mode and fixed-rate mode to introduce error control or compression ratio control. Error-bounded mode requires users to set an error bound, and fixed-rate mode means that users can set a target bitrate. 
%Error-bounded mode requires users to set an error bound, such as absolute error bound or point-wise relative error bound. The compressor ensures the differences between the original data and the reconstructed data do not exceed the user-set error-bound.
%Fixed-rate mode means that users can set a target bitrate, which is the average bits used to encode each data point, and the compressor guarantees \textcolor{black}{that} the actual bitrate of the compressed data to be lower than the user-set value.
Compared to ZFP which utilizes Discrete Fourier transform to manipulate data information, SZ predicts each data point's value by its neighboring data points in a multidimensional space with an adaptive predictor (mainly using a Lorenzo predictor~\cite{ibarria2003out}). Next, it performs an error-controlled linear-scaling quantization to convert all floating-point values to an array of integer numbers. Lastly, it performs a customized Huffman coding and lossless compression to shrink the data size significantly.
This helps SZ provide a unified error distribution between original and reconstructed data within the user-set error-bound range, which fully utilizes the error tolerance space and provides a high compression ratio \cite{di2016fast, tao2017significantly, liangerror, lu2018understanding, tao2017depth}.

SZ was first developed for CPU architectures, and has released the CUDA implementation (called cuSZ)~\cite{tian2020cusz}.
%cuSZ proposed a dual-quantization approach that first quantizes the dataset to eliminate the data dependency and then enables a fully parallel prediction. Specifically, as shown in Figure~\ref{fig:dual}, in the original SZ algorithm, the current compressed data point $q_{k-2}$ must be reconstructed before compressing the next point $d_{k-1}$, causing the data dependency issue. The dual-quantization scheme can remove the reconstruction step so that we can compress all data points independently. We refer readers to \cite{tian2020cusz} for more details about dual-quantization scheme.
Compared to lossy compression on CPUs, GPU-based lossy compression can provide much higher (de)compression throughputs~\cite{jin2020understanding}. 
cuSZ~\cite{tian2020cusz} and cuZFP~\cite{cuZFP} are existing GPU-based implementations of SZ and ZFP, respectively, which are capable of achieving tens to hundreds of GB/s (de)compression throughputs. However, these approaches are not very efficient on FPGAs, which have much lower clock frequencies compared to GPUs. Moreover, the FPGA chip space limitations prevent fitting too many instances of compression pipelines on chip. But, FPGA implementations offer several advantages over GPU implementations: \Circled{1} FPGAs can inherently provide low latency as well as deterministic latency for real-time applications. \Circled{2} FPGAs provide a high degree of user customization, and 
their implementations are easier to be integrated into other systems. 

\subsection{FPGA-based Lossy Compression}
Existing works have shown that significant performance speedups can be achieved by offloading lossy compression onto hardware. GhostSZ~\cite{xiong2019ghostsz} is the first implementation of SZ-1.0~\cite{sz16} on FPGAs. GhostSZ improves throughput by 10$\sim$85$\times$ over the SZ with a similar compression ratio and peak signal-to-noise ratio (PSNR). However, SZ-1.0 is a deprecated version which suffers from low prediction accuracy which results in low compression ratios~\cite{tian2020wavesz}.
%The prediction method used by GhostSZ may cause a significant waste of FPGA computation resources and a workload imbalance issue, as demonstrated in~\cite{tian2020wavesz}. 

waveSZ~\cite{tian2020wavesz} is another hardware implementation of SZ lossy compression. It adopts a wavefront memory layout to fit into SZ algorithm to alleviate the data dependency during the prediction process. It improves the compression ratio and throughput over GhostSZ. However, waveSZ has several drawbacks: \Circled{1} It just alleviates the data dependency using wavefront memory but does not eliminate it. As a result, its throughput does not exceed 1 GB/s. \Circled{2} Its wavefront memory layout involves rearranging data before compression, and this overhead would be relatively high when processing a large amount of data. 
\Circled{3} It only focuses on accelerating the prediction stage without handling the high overhead of Huffman coding. This, however, is the main bottleneck after fully removing the data dependency in prediction. %(will be discussed in Section \ref{subsec:challenge}).

BurstZ~\cite{sun2020burstz} is a variant of the one-dimensional ZFP algorithm and also implemented onto FPGAs. BurstZ can provide a high throughput (8 GB/s), but it suffers from a significantly lower compression ratio drop compared with the original ZFP algorithm. For example, the original ZFP algorithm achieves $21\times$ compression ratio on the NWChem dataset~\cite{NWChem} with an error bound of 0.001, whereas BurstZ only achieves $4.7\times$ compression using the same error bound. Besides, the 8GB throughput is much smaller than the throughput of current PCIe3/4 and InfiniBand. 
Moreover, DE-ZFP \cite{habboush2022zfp} is an FPGA implementation of modified ZFP by replacing the embedded encoding with a dictionary-based encoding. It focuses on maximizing throughput while minimizing the compression ratio increase. 
In addition, ZHW~\cite{ZHW} is another FPGA-based lossy compressor based on ZFP, but it is not error-bounded. So, we take BurstZ as the current state-of-the-art FPGA implementation of ZFP. % and compare CEAZ with it in our evaluation.

\subsection{MPI Collectives and MPI-I/O}

Message Passing Interface (MPI)~\cite{gropp1996high} contains two main types of operations related to parallel I/O, i.e., MPI-IO operations and collective operations. On the one hand, MPI-IO is a fundamental HPC middleware for parallel I/O. Many parallel I/O systems such as parallel HDF5~\cite{folk2011overview} and ADIOS~\cite{lofstead2008flexible} are built based on it. In MPI-IO, data is moved between files and processes by issuing read and write calls. 
The data access routines can be individual or collective. By using a collective routine, processes are coordinated with each other to optimize access to I/O devices. 
On the other hand, MPI collective operations such as scatter, gather and reduce play an important role in many HPC applications for communication. Considering that many applications use dedicated I/O nodes to periodically collect/distribute data from compute nodes and then write/read to the file system asynchronously, we also use MPI\_Gather/Scatter in this paper to evaluate our compressor in this use scenario. 
%In this paper, for demonstration purposes, we use a representative MPI collective operation (MPI\_Gather, note MPI\_scatter is a similar operation), which is used by many parallel I/Os. Since for some applications we need to specify a dedicated node as an I/O node to gather data from other nodes and then perform asynchronous file system writing.

\subsection{Research Challenges}
\label{subsec:challenge}

%\subsubsection{Huffman compression and overhead}
\noindent\paragraph{High Overhead of Huffman Coding.}
Given a set of symbols, Huffman coding generates codewords based on the evidence that not all symbols have the same probability. Instead of using fixed-length codewords, Huffman coding uses variable-length codewords based on the relative frequency of different symbols. %A fixed-length code assumes equal probability for all symbols and hence assigns the same length codewords to all symbols. In contrast to fixed length codes, Huffman coding is based on 
The principle is to use fewer bits to represent frequent symbols and more bits to represent infrequent symbols. Even though variable-length codewords can provide high compression ratio in our scenario, Huffman coding has high overhead in terms of latency, area, and power~\cite{lal20172mc}. To achieve a high overall compression throughput, certain key challenges in Huffman coding need be addressed.

\begin{figure}[t]
    \centering
    % \vspace{4mm}
        \includegraphics[width=1.0\linewidth,
    trim={7.4in 8.1in 7.2in 8.1in}, clip
    ]{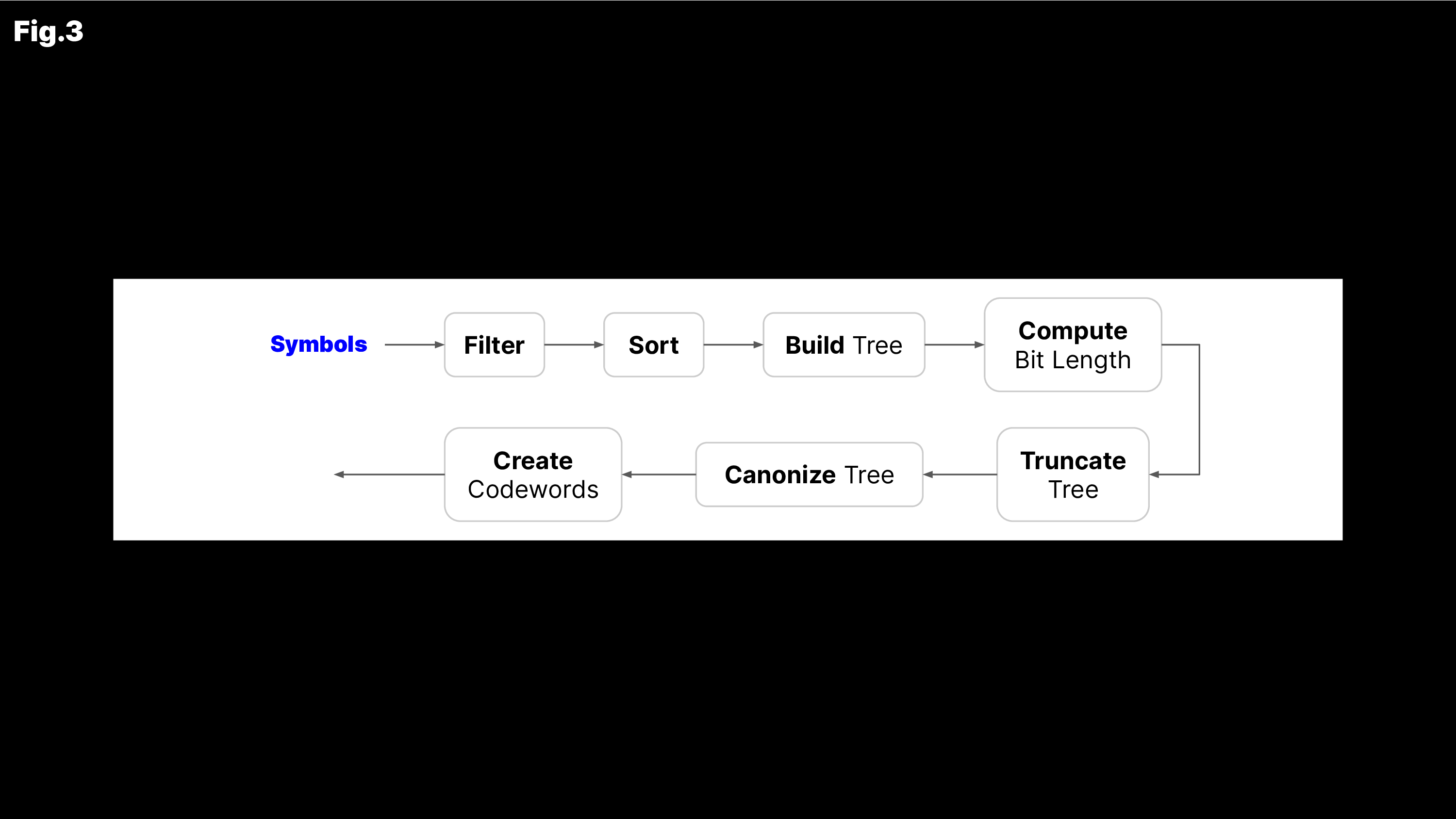}
%    \vspace{-2mm}
    \caption{The process of canonical Huffman encoding.}
    \label{fig:Huffman}
%    \vspace{-2mm}
\end{figure}

\noindent\paragraph{Challenge of Codewords Generation.} The first challenge is to build a Huffman tree and generate codewords within limited hardware clock cycles to meet high-throughput requirements. Our goal is to accelerate MPI collective I/O in real time through compression. So we hope to reduce the compression latency as much as possible.
In addition, generating codewords needs 7 steps: filter, sort, create tree, compute bit length, truncate tree, canonize tree, and create codewords, as shown in Figure~\ref{fig:Huffman}. This procedure is a serial process that is hard to be parallelized on FPGAs or GPUs. We make full use of the characteristics of the FPGAs to speed up this process by pipelining. However, the latency presented in Figure~\ref{fig:latency} is still large.

\paragraph{Challenge of Predefined Codewords.}
Inspired by ~\cite{lal20172mc}, we will use predefined codewords at the beginning and update the codewords during the runtime. This method introduces the second challenge: how to generate suitable codewords to cover the features of all the scientific datasets?

\section{Design Methodology}
\label{sec:design}
In this section, we describe our proposed FPGA-based lossy compressor CEAZ and parallel I/O accelerator. Specifically, we will first overview the design of CEAZ and then describe our efficient and adaptive Huffman coder. Finally, we describe our proposed parallel I/O accelerator integrated with CEAZ. 

\begin{figure}[h]
    \centering
%    \vspace{-1mm}
    \includegraphics[width=0.8\linewidth]{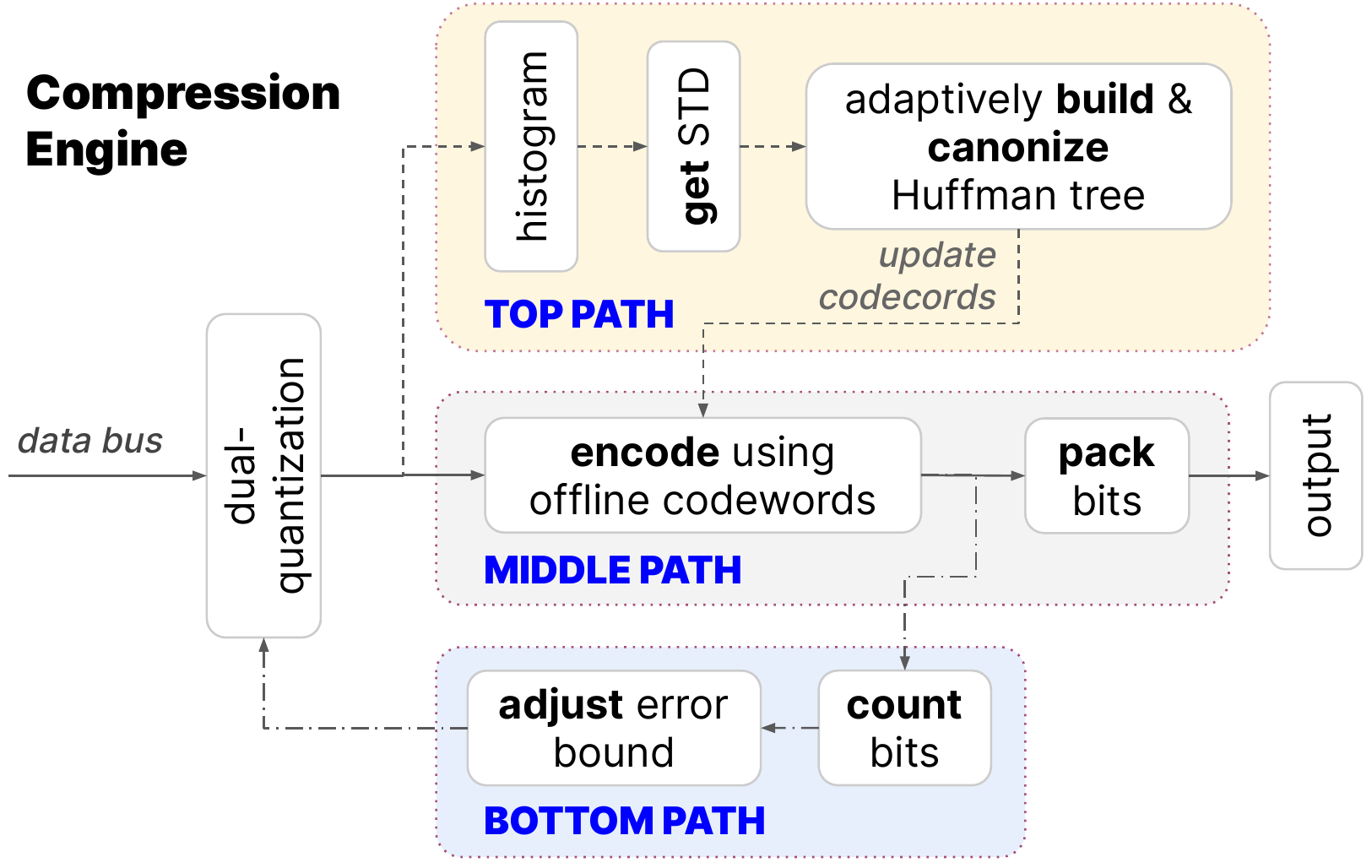}
    \caption{Design of our proposed lossy compression engine CEAZ.}
    \label{fig:engine}
%    \vspace{-1mm}
\end{figure}

\subsection{Overview of CEAZ Engine}

We show our proposed lossy compression engine in Figure~\ref{fig:engine}. It has three main dataflow paths. 
On the top dataflow path, we preprocess float-point data using a dual quantization algorithm \cite{tian2020cusz} (abbreviated as dual-quant), which generates integers as symbols (quantization codes) for the following Huffman coding. We collect frequencies of symbols using a histogram and calculate the standard deviation (STD) of frequencies. According to the STD value, we will decide whether to build a new Huffman tree based on the symbol frequencies or not (will be discussed in the next section). 

\begin{figure}[b]
    \centering
%    \vspace{1mm}
%    \includegraphics[width=0.9\linewidth]{figures/pipeline.pdf}
    \includegraphics[width=1.0\linewidth,
    trim={3.5in 6.1in 3.5in 6.1in}, clip]{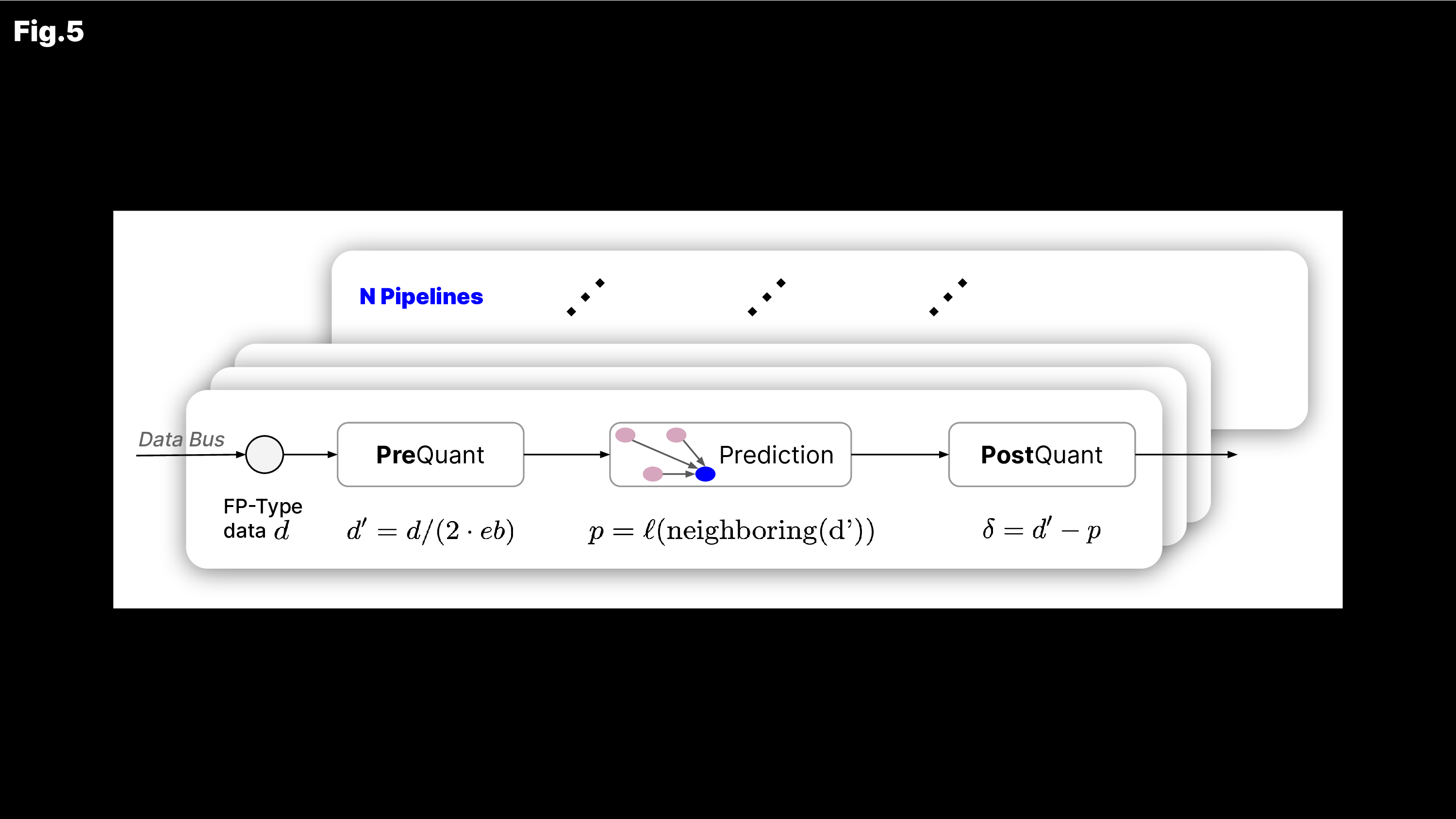}
    \caption{Design of our adopted dual-quantization pipeline.}
    \label{fig:pipeline}
%    \vspace{-1mm}
\end{figure}

Figure~\ref{fig:pipeline} shows our dual-quant pipeline design. Dual-quant is a novel two-phase prediction-quantization approach, which can completely eliminate the data dependency in the prediction and quantization steps. Our dual-quant consists of two steps: \textcolor{black}{prequantization} and postquantization. \textcolor{black}{Given a float point data $d$, we first quantize it based on the user-set error bound and convert it to an integer data $d'$.} \textcolor{black}{After the prequantization}, we can calculate its predicted value based on its neighboring values (denoted $\text{neighboring}(d')$) using Lorenzo predictor (denoted $\ell$), $p = \ell(\text{neighboring}(d'))$. The second step, called postquantization, computes the difference $\delta$ between the predicted value and the prequantized value. $\delta$ will be compressed by Huffman compression. Since the dual-quant part has no data dependency, we could instantiate $N$ pipelines to process $N$ float-point data in parallel.
On the middle dataflow path, we directly encode symbols using existing codewords for seeking high throughput. The encoder can find the codeword corresponding to each symbol and output it. We then pack variable-length encoded (compressed) symbols to save the storage space.
On the bottom dataflow path, we feed back total bits of encoded symbols to estimate compression ratio, and then adjust the error bound. 

Our proposed compression engine has two working modes: fixed accuracy (i.e., error bounded) and fixed ratio (i.e., fixed bit-rate). The fixed-accuracy mode ensures information loss of compressed data is within the specific error bound, whereas the fixed-ratio mode ensures the transfer of compressed data has a consistent throughput.
For the fixed-accuracy mode, we need to define an upper bound of errors that can be tolerated by applications and keep using this error bound through the whole compression process. 
For the fixed-ratio mode, we can set a suitable error bound to achieve a target compression ratio, as a higher error bound leads to a higher compression ratio.
Specifically, we adjust the error bound as follows: 
\begin{enumerate}[noitemsep, topsep=3pt, leftmargin=1.3em]
    \item We use $C = \frac{\text{\ttfamily TotalBits(original data)}}{\text{\ttfamily TotalBits(compressed data)}}$ to estimate the compression ratio, where $\text{\ttfamily TotalBits(original data)} = W * N$. Here $W$ is the bit-rate of original data; for single/double floating-point data, $W$ is 32/64 bits per value; $N$ is the total number of data points that have been compressed.
    \item We calculate the compressed and target bit-rates by $B = \frac{W}{C}$ and $B_{\text{target}} = \frac{W}{C_{\text{target}}}$, respectively.
    \item We adjust the error bound by Equation (\ref{equ-2}) (will be discussed in Section \ref{subsec:offline}).
\end{enumerate}
%\Circled{1} We estimate the compression ratio by $C = \frac{\text{\ttfamily TotalBits(original data)}}{\text{\ttfamily TotalBits(compressed data)}}$, where $\text{TotalBits(original data)} = W * N$. Here $W$ is the bit-rate of original data; for single/double floating-point data, $W$ is 32/64 bits per value; $N$ is the total number of data points that have been compressed. \Circled{2} We calculate the compressed and target bit-rates by $B = \frac{W}{C}$ and $B_{\text{target}} = \frac{W}{C_{\text{target}}}$, respectively. \Circled{3} We adjust the error bound by Equation (\ref{equ-2}) (will be discussed in Section \ref{subsec:offline}).

\subsection{Efficient and Adaptive Huffman Coder}
\begin{figure}[t]
    \centering
%    \vspace{-1mm}
    \includegraphics[width=\linewidth]{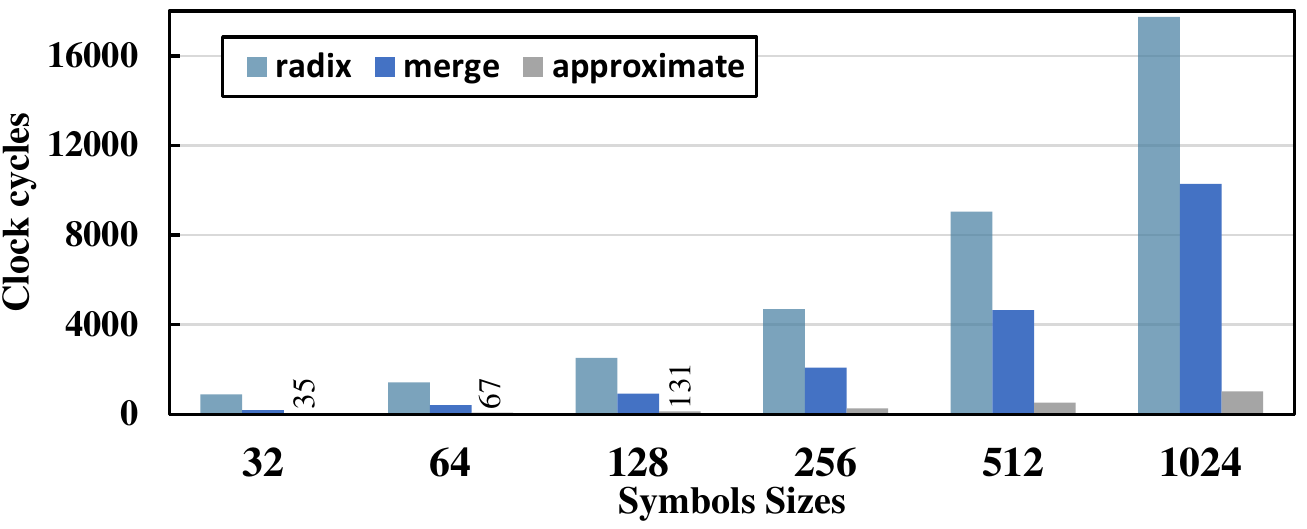}
    \caption{Latencies of different sorts with different symbol sizes.}
    \label{fig:latency}
%    \vspace{-2mm}
\end{figure}

%\subsubsection{Fast approximate sort based on Lorenzo predictor's feature}
\subsubsection{Fast approximate sort}
In order to build the Huffman tree, we must sort the symbols based on their frequencies after we filter out symbols with a frequency of zero. 
Previous works~\cite{kastner2018parallel, Vitis_Libraries} use radix sort to reduce the utilization of hardware resources. However, we find that radix sort typically takes more than 30\% of the total time when we break down the execution time of generating codewords. This is because the time complexity of radix sort is $O(d\times(n+b))$ (where $b$ is the base to represent numbers and $d$ is the maximum number of digits) with $d = 32$ and $b = 10$ typically. 
% Inspired by ~\cite{kastner2018parallel}, we replace radix sort with merge sort. This is because merge sort has a lower time complexity (i.e., $O(n\log n)$) than radix sort (i.e., $O(d\times(n+b))$) considering $\log n$ is relatively small such as 10 (lower than $d = 32$) in our case. 
We also adopt the non-recursive way of merge sort in our hardware implementation. However, we identify that Vitis HLS requires the size of the array to be sorted is a constant and a power of two~\cite{Vivado-ug}, which prevents us from using the original merge-sort hardware implementation. This is because the non-zero frequencies to be sorted in our case are neither fixed nor the power of two.

\begin{figure}[h]
\centering
% \vspace{-4mm}
\begin{subfigure}{0.49\linewidth}\centering
    % \vspace{-3mm}
    \includegraphics[width=0.99\linewidth]{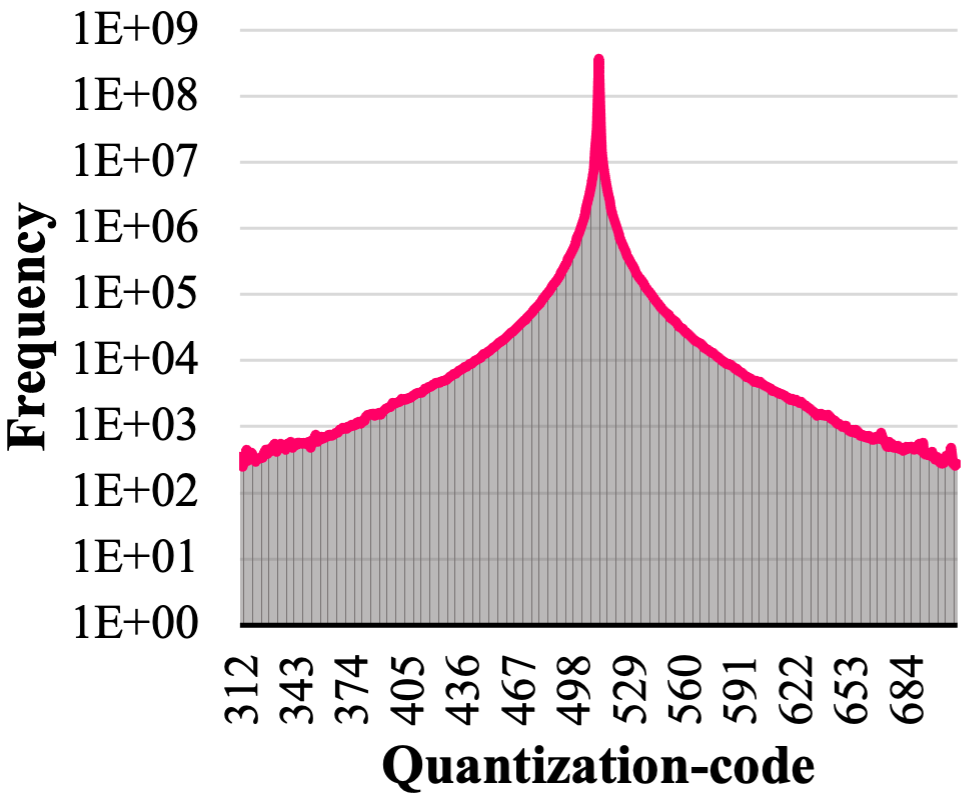}
%    \vspace{-3mm}
	\caption{NWChem}\label{fig:Freq_NWChe}
%	\vspace{2mm}
\end{subfigure}
\begin{subfigure}{0.49\linewidth}\centering
    \includegraphics[width=0.99\linewidth]{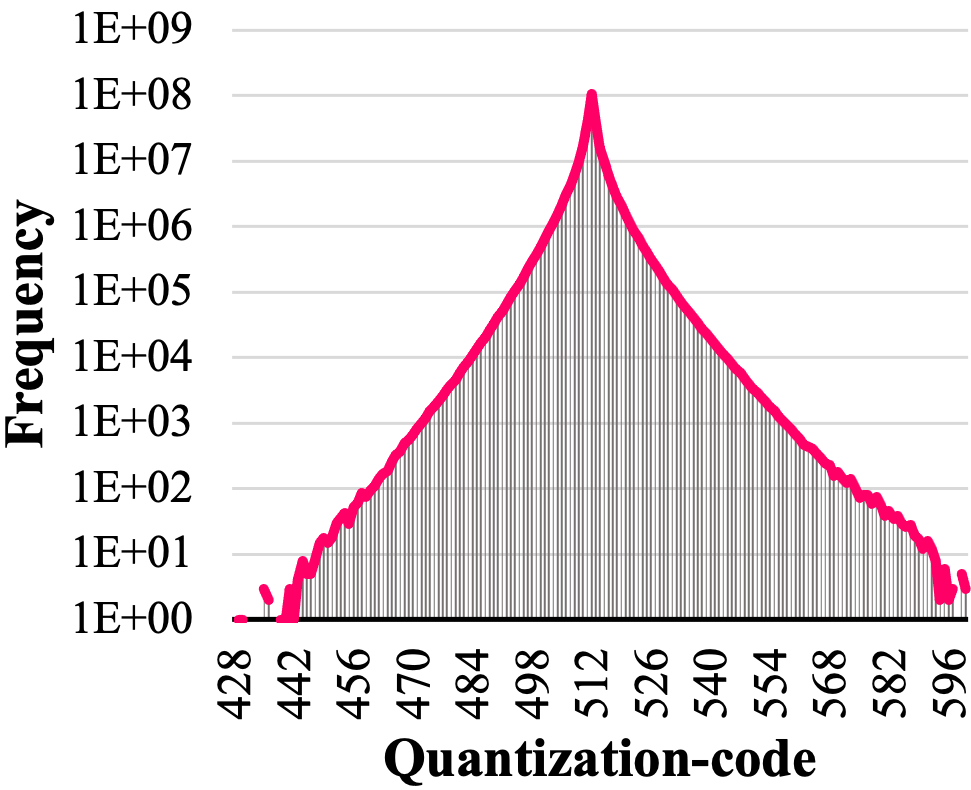}
%    \vspace{-3mm}
	\caption{HACC}\label{fig:Freq_HACC}
%	\vspace{2mm}
\end{subfigure}
\begin{subfigure}{0.49\linewidth}\centering
    \includegraphics[width=0.99\linewidth]{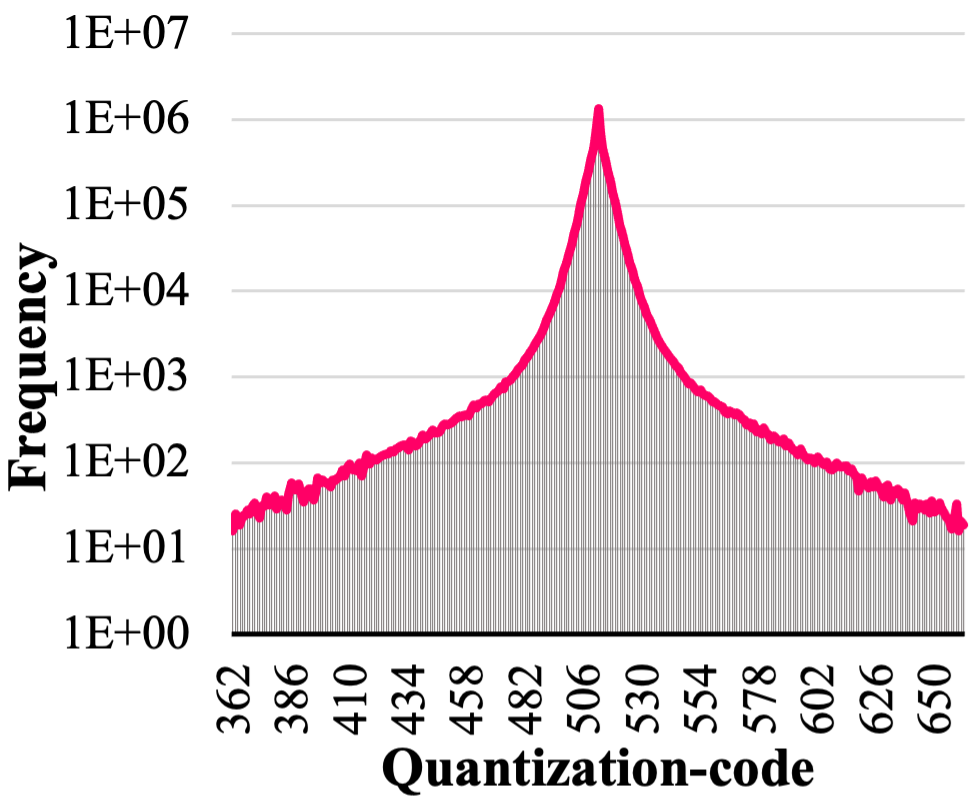}
%    \vspace{-3mm}
	\caption{CESM}\label{fig:Freq_CESM}
%	\vspace{2mm}
\end{subfigure}
\begin{subfigure}{0.49\linewidth}\centering
    % \vspace{-3mm}
    \includegraphics[width=0.99\linewidth]{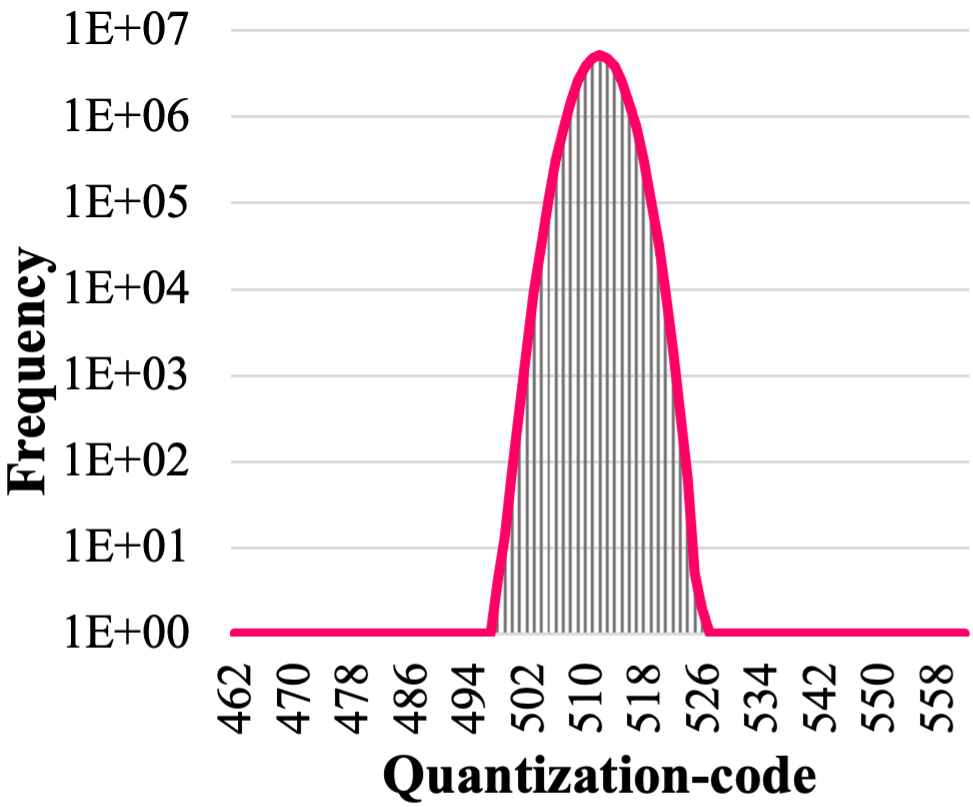}
%    \vspace{-3mm}
	\caption{Brown}\label{fig:Freq_Brown}
%	\vspace{2mm}
\end{subfigure}
\caption{Distribution of symbol frequencies on four scientific datasets, i.e., NWChem~\cite{NWChem}, HACC~\cite{HACC}, CESM~\cite{cesm-atm}, and Brown~\cite{Brown-Samples}.}
\label{fig:distribution}
\end{figure}

We note that the frequencies of symbols that are generated by Lorenzo predictor and linear-scaling quantization~\cite{tao2017significantly} are symmetric, as shown in prior studies~\cite{jin2021adaptive}. We verify this in our experiments with 1024 symbols, as shown in Figure~\ref{fig:distribution} (symmetric with respect to symbol 513). This feature inspires us to use an approximate sort to improve the efficiency, since Huffman coding can accept the approximately sorted symbols, which would not notably degrade the compression ratio. 
Specifically, assuming $A$ is an unsorted array with symbols, and $O$ stores sorted symbols. We initialize two indexes $l$ and $h$ to represent the symbols to the left and right of the middle symbol. We compare the frequencies of the two symbols and store them in a correct order into $O$. We then decrease $l$ by 1 and increase $h$ by 1 and repeat this till all symbols are sorted.

We describe our proposed approximate sort algorithm in Algorithm \ref{alg:sort} in detail. It has the time complexity of $O(\frac{n}{2})$, which is lower than the time complexity of radix sort and merge sort. We compare sort time of different sort algorithms with different symbol sizes. The results shown in Figure~\ref{fig:latency} illustrate that our approximate sort saves the sort time by 7.5$\times$ on average over the merge sort. More details about our evaluation platform will be shown in Section \ref{sec:evaluation}.

\setlength{\textfloatsep}{0pt}
\begin{algorithm}[t]
\footnotesize\sffamily
\normalfont
\SetAlgoLined
\setcounter{AlgoLine}{0}
\KwResult{A approximately sorted array}
\BlankLine
$S$: defined structure contains two members: symbol, and its frequency\par
$A$: input array, its data type is $S$, $len$: length of input array, $i$: index of $A$\par
$p$: index of symbol 513 in $A$, $m$: index of the midpoint of $A$, $l$: index, $h$: index\par
$O$: output sorted array, $j$: index of $O$\par
$t$: loop count\par
% \BlankLine
$l = p - 1$, $h = p + 1$, $j = len - 2$, $O[len-1] = A[p]$ \par
\eIf{$p \leq m$}
 {
  $t = p$
 }
 {
  $t = len - p - 1$
 }

\For{$i \gets 1$ to $rows$}
{
  Count
 \eIf{$A[l].frequency \leq A[h].frequency$}
  {
  $O[j] = A[h]$\par
  $O[j-1] = A[l]$
  }
  {
  $O[j] = A[l]$\par
  $O[j-1] = A[h]$
  }
$l = l - 1$, $h = h + 1$, $j = j - 2$\par
}
CopyRemaining($A$, $O$) /* copy remaining data from $A$ to $O$ */ \par
\caption{Proposed fast sort based on Lorenzo predictor's feature.}
\label{alg:sort}
\end{algorithm}

\subsubsection{Offline Huffman codewords generation}
\label{subsec:offline}
On the premise of meeting the acceptable reduction in compression ratio, we propose to combine offline and online Huffman codewords generation strategies in order to improve the throughput as much as possible. As shown in figure~\ref{fig:engine}, the symbols generated at the beginning by dual-quant will be encoded by offline codewords directly; at the same time, we also collect the frequencies of symbols. We will generate new Huffman codewords if the change of STD of symbol frequencies is greater than the threshold $\tau$. $\tau$ is a hyper-parameter, and we will discuss it in the next section. 

In order to make offline codewords representative, we generate corresponding offline codewords for various types of datasets that are currently available.
The current types of offline codewords include Climate, Cosmology, Molecular, and Physics.
When we target to compress a certain type of dataset, we will use the offline codewords corresponding to this type. 
However, when we encounter a new type of dataset, we will first use the average offline codewords. Then we will add the offline codewords of this new type into our offline codewords repository for future uses.

Specifically, we generate offline codewords based on the following four steps: \Circled{1} We set a suitable error bound to let our compressor have a similar compression ratio on different datasets under the same type. \Circled{2} We collect symbol frequencies on different datasets under the same type. \Circled{3} We calculate the average symbol frequencies from collected frequencies. The average symbol frequencies are used to generate offline codewords. \Circled{4} We store the offline codewords of this type into our offline codewords repository. In order to make offline codewords representative and promising for high compression ratio, we collect symbol frequencies based on all the real-world datasets from the Scientific Data Reduction Benchmarks (SDRBench) \cite{sdrbench} and Datasets for Benchmarking Floating-Point Compressors \cite{knorr2020datasets}. Figure~\ref{fig:offline} shows the ratio degradation by using the offline codewords (will be showed in Section \ref{subsec:evaluation-update}).

Moreover, to generate the average offline codewords, we use the following three steps: \Circled{1} We set a suitable error bound to let our compressor have a similar compression ratio on different datasets under different types. \Circled{2} We collect symbol frequencies on different datasets under different types. \Circled{3} We calculate the average symbol frequencies from collected frequencies. The average symbol frequencies are used to generate the average offline codewords.

Using different error bounds to compress the same dataset results in different histograms of symbols (i.e., different distributions of symbol frequencies). For example, using a larger error bound results in a tighter histogram of symbols compared to using a smaller error bound. In extreme cases where very large error bounds are used, there can be only a few symbols for Huffman coding. To make the offline codewords adaptive for a wide range of datasets, we must choose suitable error bounds for multiple scientific datasets, which can result in a similar histogram of symbols after employing the Lorenzo predictor. In other words, we must control the error bound for each dataset to provide a similar ratio. 
Instead of using a trial-and-error approach to search the suitable error bound for every dataset, we provide a theoretical analysis to predict the error bound given a target ratio based on one-time sampling.

A naive solution to align different datasets with a similar compression ratio is to use the same value-ranged-based relative error bound\footnote{Note that unlike the pointwise relative error that is compared with each data value, value-range-based relative error is compared with value range.} instead of the same absolute error bound.
While using the same value-ranged-based relative error bound for different datasets can reduce the divergence of their symbols' histograms, it cannot guarantee that the compression ratio of different datasets is similar to each other.
In our experiment, we identify the compression ratio range of $4$$\sim$$13\times$ when using the same value-ranged-based relative error bound for multiple scientific datasets. 
Our proposed solution considers the efficiency of Huffman coding affected by error bound to accurately estimate the error bound for a target compression ratio. We assume the bit-rate of symbol after Huffman encoding is:
\begin{align}
    \text{mean}(L) = \sum_{i=0}^{n}P(s_i)L(s_i) \approx \sum_{i=0}^{n}P(s_i)\log_2P(s_i),
    \label{equ-1} 
\end{align}
where $n$ is the number of different Huffman code, $P$ is the probability of given code $s_i$, $L$ is the length of given code $s_i$. We further represent the Huffman code length based on its probability with binary base-2 numeral system. Note that in our case, 1024 symbols are used for Huffman coding and thus are sufficient for this simplification. Consider a given error bound $eb$ can provide a bit-rate of $B$, when doubling the error bound to $2eb$, the symbols' histogram also shrinks accordingly where the total number of symbols is reduced by $2\times$ and the possibility of each symbol is increased by $2\times$. In this case, the bit-rate should be:
\begin{align}
    B' & = \sum_{i=0}^{n/2}P'(s_i)\log_2P'(s_i) \approx \sum_{i=0}^{n/2}(P(s_{2i-1})\log_2P'(s_{2i-1}) \notag \\ 
        & + P(s_{2i})\log_2P'(s_{2i}))-1 = B-1 
    \label{equ-2} 
\end{align}

Thus, we conclude that by doubling the error bound, the bit-rate should increase by $1$. Furthermore, we can derive that if the compression bit-rate is $B$ under the error bound $eb$, then under the new error bound $Neb$ the predicted bit-rate is $B'=B-\log_2N$.
Note that the SZ algorithm uses previous data points' quantized values to predict the value of current point based on Lorenzo prediction, which means different error bounds would affect the shape of symbols' histogram. However, based on our experiments, this only applies to very large error bounds and hence few quantization bins. In our case, we simplify this to a fixed symbols' histogram shape under different error bounds, yielding a precise $2\times$ shrink when doubling the error bound.

With the above analysis, we can simply compress each scientific dataset once with the same value-ranged-based relative error bound $eb$ and compute the optimized error bound $eb'$ for the target bit-rate $B_{\text{target}}$ based on the current bit-rate $B$ by %the following equation:
%\begin{align}
$eb' = 2^{B - B_{\text{target}}}eb$.
%    \label{equ-3} 
%\end{align}

\subsubsection{Adaptive online codewords updates}
In general, on one hand, the more frequently we update the codewords, the closer we can get to the optimal codewords in terms of compression ratio. On the other hand, too frequently updating codewords may decrease the throughput, since the newly generated codewords need to be stored. Moreover, if we do not update the codewords, the compression ratio may decrease as well, because the old codewords are too outdated to reflect current distribution of symbol frequencies. In order to solve this problem, we use two effective metrics to determine when to generate new codewords: \Circled{1} the storage overhead of codewords and \Circled{2} the change of distribution of symbol frequencies.

For example, suppose we have $S$ symbols and $S$ codewords. Each codeword is $B$ bits on average after canonization process. We have $\text{\ttfamily size(codewords)} = S \times B$. Our target compression ratio is $C$. The bit-rate of original data is $W$. For single or double floating-point data set, the bit width is 32 or 64 bits per value. The bit-rate $R$ is $\frac{W}{C}$. Bit-rate $R$ can also be regarded as an average bit length of compressed data. We have $\text{\ttfamily size(compressed data)} = R * N$, where $N$ is the total number of original data. Assume the ratio of the codewords size to the compressed data size is $O$, 
%i.e., $O =\frac{\text{\ttfamily size(codewords)}}{\text{\ttfamily size(compressed data)}}$,
if the codewords overhead is set to be less than 10\%, $\frac{S \times B}{S \times B + C \times N} \leq 10\%$. 
% For example, we have 1k (k=1024) symbols, and each codeword consumes 8 bits. If we set the target compression ratio to be 10, then we have $N > 24k$.

The symbols generated by dual-quant present a centralized and symmetric distribution, as shown in Figure~\ref{fig:distribution}. The generated codewords are highly related to the distribution of symbol frequencies~\cite{blanes2019lower}. These good characteristics inspire us to evaluate the similarity of two sets of symbol frequencies using STD. Specifically, assume $\sigma_0$ is the STD obtained from the previous data chunk, $\sigma_1$ is the STD obtained from the current data chunk, and $\chi = \lvert\sigma_0 - \sigma_1\rvert$. We define a set of thresholds $\tau_x$ and propose the following strategy: 
\begin{itemize}[noitemsep, topsep=3pt, leftmargin=1.3em]
    \item We will not generate new codewords if $\chi \leq \tau_0$ (two frequencies with similar distributions generate almost identical codewords) but keep using the old codewords;
    \item We will generate new codewords if $\tau_0 < \chi \leq \tau_1$;
    \item We will use the offline Huffman codewords if $\chi \geq \tau_1$.
\end{itemize}

We are processing data that is completely different from the previous data if distribution changes drastically. We need to clear the histogram of compression engine and collect new symbol frequencies. We set $\tau_0$ and $\tau_1$ as 3.05 and 4.88, respectively, after comprehensive experiments (will be discussed in detail in Section \ref{sec:eval-std}).

Note that our design is different from other Huffman coding works in terms of adaptivity. For example, Tian et al. \cite{tian2021revisiting} proposed a reduction-based scheme for GPUs that iteratively merges the encoded symbols and adaptively determines the number of merge iterations. However, CEAZ only builds a new codebook for the data chunk when the change of its histogram exceeds a threshold in order to target FPGA with limited resources and low clock frequency.
%Moreover, to reduce the overhead of building Huffman codebook, Tian’s work proposed a parallel method exploiting GPU parallelism, whereas our solution proposes to use an efficient approximate sort under limited resources and low clock frequency for FPGA. 
%Our work is completely different from Tian’s work regarding the adaptivity of Huffman coding. Tian’s work focused on the encoding stage in Huffman coding. It proposed a reduction-based encoding scheme that iteratively merges the encoded symbols and adaptively determines the number of merge iterations. However, our design focuses on the codewords generation stage in Huffman coding. CEAZ only builds a new codebook for the data chunk when its histogram’s change exceeds a threshold (but Tian’s work always builds a new Huffman codebook for every data chunk).

\subsection{Parallel I/O Accelerator}

\begin{figure}[h]
    \centering
%    \vspace{-1mm}
    \includegraphics[width=\linewidth,
    trim={5.5in 8in 5.5in 8in}, clip]{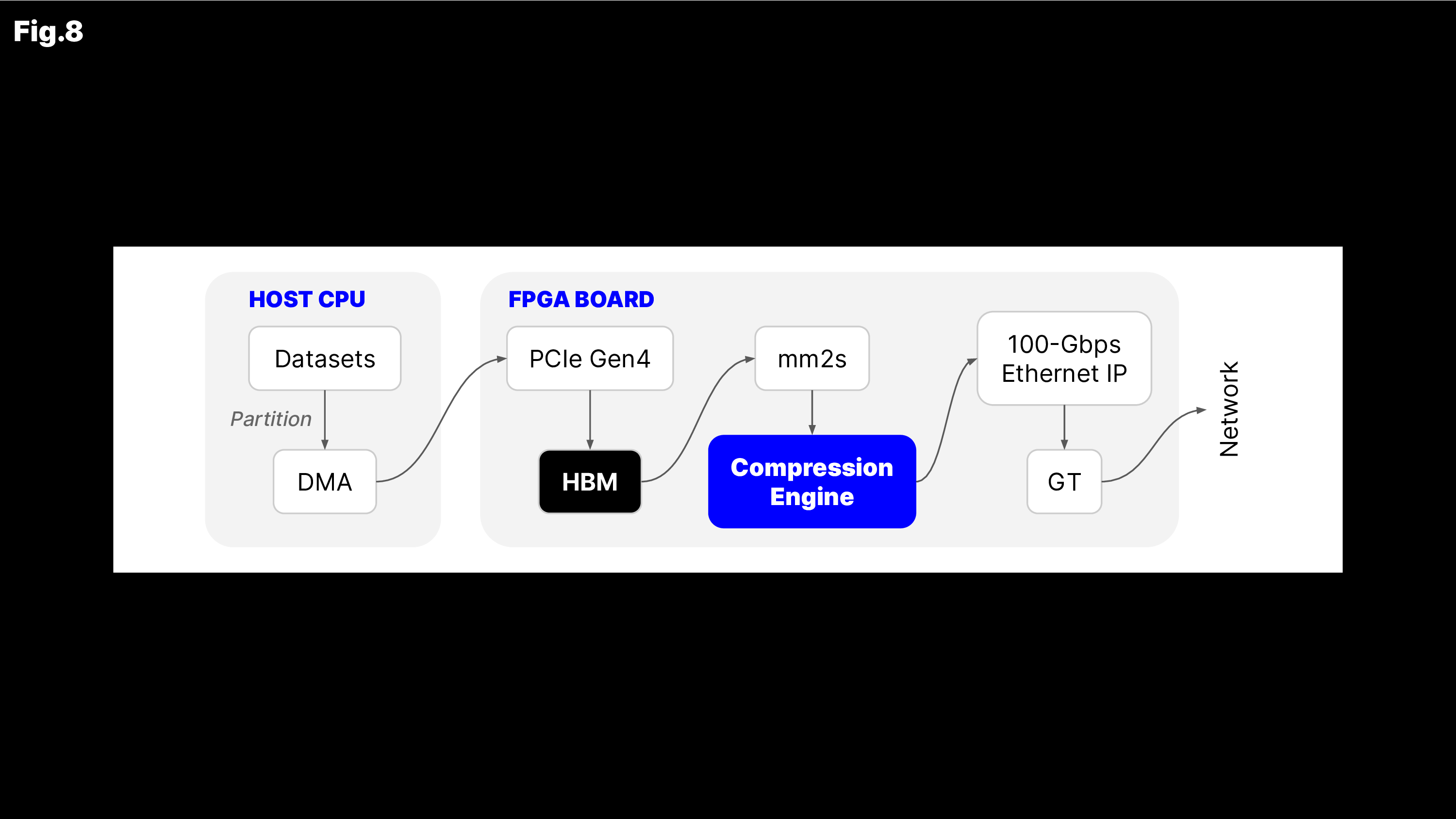}
 %   \vspace{-1mm}
    \caption{Overview of system architecture integrated with CEAZ.}
    \label{fig:NIC}
    \vspace{2mm}
\end{figure}

Figure~\ref{fig:NIC} shows the overview of our system architecture integrated with CEAZ compression engine. Our system includes two parts: 
\Circled{1} The host partitions the input dataset and feed the chucked data through the PCIe. Raw data is buffered in high-bandwidth memory (HBM) with 460 GB/s bandwidth and converted into stream data by memory to stream (mm2s) unit. 
\Circled{2} The compression engine compresses the stream data in real-time and output compressed data. Ethernet intellectual property (IP) packs the compressed data according to the network protocol. QSFP28 (fiber optical transceiver) gigabit transceiver (GT) finally outputs packed data into network. 

\begin{figure}[H]
    \centering
    % \vspace{1mm}
    \includegraphics[width=.9\linewidth]{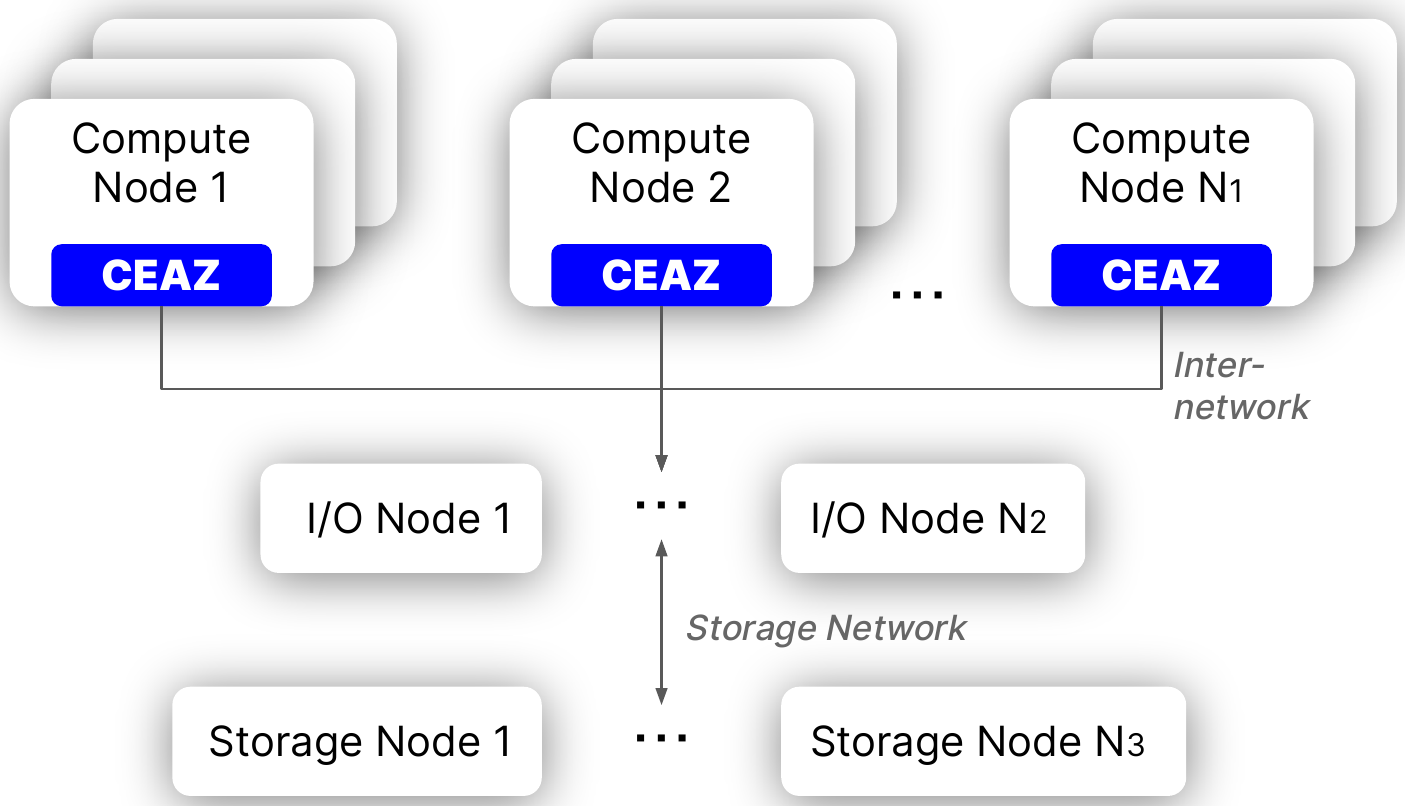}
    \caption{Overview of CEAZ-supported parallel I/O system.}
    \label{fig:IO}
%    \vspace{2mm}
\end{figure}

Many scientific applications, such as cosmology simulations, need to periodically dump a huge amount of raw simulation data to the storage for post-hoc analysis and visualization after simulations. Data across all computing nodes needs to be aggregated to the storage node(s). Even though state-of-the-art supercomputers are using InifniBand interconnect (e.g., 200 Gb/s), it can take hours to complete the data aggregation and save (e.g., 1.5 TB/s of aggregated I/O bandwidth and 4.85 PB memory capacity in Fugaku supercomputer \cite{nakao2020performance}). 
Therefore, we propose to apply CEAZ to the future HPC systems, as shown in Figure~\ref{fig:IO}. 

Specifically, CEAZ is directly integrated into the FPGA-based SmartNIC
%\footnote{FPGA-based SmartNIC is a network interface card (network adapter) that offloads processing tasks that the system CPU would normally handle. Using its own on-board FPGAs, the FPGA-based SmartNIC can perform any combination of networking tasks.}
in each computing node and used to compress the raw data before transmitting it to the storage system via interconnection network. 
%Deployment of CEAZ. U280 will perform as SmartNIC, Xilinx Alveo SN1000 \cite{sn1000} is the industry’s first SmartNIC offering software-defined hardware acceleration for all function offloads in a single platform. The FPGA chips of SN1000 and U280 are the same type (Xilinx Virtex Ultrascale+). Beside, CEAZ just consumes 26K LUT, which can be fitted into mainstream SmartNIC. There is no change to storage system since data is compressed before sent into network. For storage system, the compressed data is just like normal data.
There are two main scenarios to use the CEAZ-compressed data in the storage: \Circled{1} checkpoint/restart and \Circled{2} post analysis and visualization.
% (1) Checkpoint/restart: to restart the application, the system needs to load the compressed checkpoint from the storage back to each computing node through the FPGA-based SmartNIC, and the decompression engine of CEAZ will reconstruct the whole dataset/checkpoint and pass it to the CPU memories for the application. 
% (2) Post analysis and visualization: users can use the CPU version of our lossy compressor to only decompress the data needed for analysis and visualization, instead of loading the whole compressed data back and relying on the decompression engine in CEAZ.
Thus, there is no need to change the storage system to adapt to our design, since the data is compressed (decompressed) before (after) sending (receiving) to (from) the network adapter. 
Note that similar to the FPGA-based SmartNIC, the emerging Data Processing Unit (DPUs) \cite{DPU}---a class of programmable processor---based SmartNIC can also offload and improve application performance for communications and storage. We will extend CEAZ to DPU-based systems in future work.

\section{Experimental Evaluation}
\label{sec:evaluation}
In this section, we evaluate our proposed FPGA-based error-bounded lossy compressor CEAZ and demonstrate its effectiveness in two perspectives: (1) the effectiveness of the proposed adaptive compression algorithm and the performance (i.e., throughput and latency) of its accelerator implementation, and (2) the improvement of parallel I/O supported by CEAZ in different scales. 

\subsection{Experimental Setup}
\noindent\paragraph{Experimental Platform}
We use two platforms as our testbed. The first platform is Xilinx Alveo U280 Data Center accelerator card, which is equipped with a PCIe Gen4x8 with CCIX to leverage the latest server interconnect infrastructure for high-bandwidth host processors, 8 GB HBM2 and 32 GB on-board DDR4 DRAM. CEAZ is implemented with Xilinx Vitis unified software platform (v.2020.2) \cite{vitis}. The second platform (for parallel I/Os) is Summit~\cite{Summit}, which is one of the most powerful supercomputers in the world. We perform GPU experiments on an NVIDIA Tesla V100 GPU. 

%In order to compare the throughput with compressors on CPU and GPU. The second platform is Bridges-2~\cite{Bridges-2}, which is the newest supercomputer at Pittsburgh Supercomputing Center (PSC). Bridges-2 has 504 nodes (each has two 64-core AMD EPYC 7742 CPUs and 256+ GB RAM), Mellanox ConnectX-6 HDR InfinBband 200Gb/s Adapter, and 15 PB Lustre parallel file system (with 142 GB/s storage bandwidth) \cite{Bridges-2-storage}. Bridges-2 also has 24 GPU nodes, and each is equipped with eight NVIDIA SXM2 32 GB V100 GPUs. We perform our GPU experiments on one GPU node.

\noindent\paragraph{Test Datasets}
To conduct our evaluation and comparison under realistic scenarios, we use six real-world datasets from the Scientific Data Reduction Benchmarks~\cite{sdrbench}. Datasets belong to various domains: \Circled{1} 2D CESM-ATM climate simulation~\cite{cesm-atm}. \Circled{2} 1D HACC cosmology particle simulation~\cite{HACC}. \Circled{3} 3D NYX adaptive mesh hydrodynamics and N-body cosmological simulation ~\cite{HACC}. \Circled{4} 1D NWChem two-electron repulsion integrals computed over Gaussian-type orbital basis sets~\cite{NWChem}. \textcolor{black}{The sizes of the three fields are 102,953,248, 801,098,891, and 712,996,037, respectively.} \Circled{5} 3D S3D Combustion simulation~\cite{S3D}. \textcolor{black}{Each raw file needs to be split into 11 files to form the data size of $500\times500\times500$}. \Circled{6} 1D Brown Samples synthetic and generated to specified regularity~\cite{Brown-Samples}. More details about the datasets can be found in Table~\ref{tab:Datasets}.

\begin{table}[t]
%    \vspace{-1mm}
    \caption{Test datasets from Scientific Data Reduction Benchmarks.}
    \centering
    \resizebox{.95\linewidth}{!}{\begin{tabular}{@{} >{\bfseries}l @{}rrrrr@{}}
\toprule
	\textbf{\sffamily type}
	& \textbf{\sffamily name} 
    & \textbf{\sffamily \# fields} 
    & \sffamily\bfseries precision
    & \textbf{\sffamily dimensions}
    & \textbf{\sffamily size}
\\
\midrule

\multirow{1}{*}{Climate} & \sffamily CESM
    & 77 
    & \ttfamily float
    & 1,800$\times$3,600 
    & 1.86 GB
\\
% & \sffamily Hurricane
% 	& 13 
% 	& \ttfamily float
% 	& 100$\times$500$\times$500
% 	& 1.25 GB
% \\
\midrule
\multirow{2}{*}{Cosmology} & \sffamily HACC
	& 6 
	& \ttfamily float
	& 280,953,867
	& 6.28 GB
\\
& \sffamily NYX
    & 6 
    & \ttfamily float
    & 512$\times$512$\times$512
    & 3.0 GB
\\
\midrule
\multirow{1}{*}{Molecular} & \sffamily NWChem 
    & 3 
    & \ttfamily double 
    & 1,617,048,176
    & 12.05 GB
\\
\midrule
\multirow{1}{*}{Physics} & \sffamily S3D 
    & 55 
    & \ttfamily double 
    & 500$\times$500$\times$500 
    & 51.22 GB
\\
\midrule
\multirow{1}{*}{Other} & \sffamily Brown 
    & 3 
    & \ttfamily double 
    & 33,554,433
    & 0.75 GB
\\
\bottomrule
\end{tabular}}
    \vspace{1mm}
    \label{tab:Datasets}
\end{table}

\subsection{Resource Utilization and Clock Frequency}
Table~\ref{tab:utilization} shows the breakdown of hardware resource utilization of CEAZ. To provide a fair comparison with BurstZ, we implement 32 pipelines for single-precision datasets and 16 pipelines for double-precision datasets. U280 has 2 HBM2 stacks and each stack has 16 channels. Each channel is 32-bits data width. 32 pipelines require a data width of exactly 1024. The BRAM\_18K is dual-port RAM module instantiated into the FPGA for on-chip storage, and its size is 18k bits. DSP block is an arithmetic logic unit. FF represents for flip-flop. Look-up table (LUT) is the basic building block of an FPGA and is capable of implementing any logic function. The running frequency based on our measurement is slightly lower than the clock frequency we set (i.e., 300 MHz) but is still around 300 MHz. Note that N/A means the utilization data is not provided in the BurstZ work \cite{sun2020burstz}.

\begin{table}[t]
%    \vspace{-1mm}
    \caption{Hardware resource utilization.}
    \centering\small
    \resizebox{.9\linewidth}{!}{\begin{tabular}{@{} >{\bfseries}l cccccc @{}}
\toprule
  &board &BRAM\_18K &DSP &FF &LUT &percent \\
\midrule
BurstZ &VCU118 &222 &N/A &N/A &125000 &<5\% \\
CEAZ   &U280 &475 &256 &67623 &302407 &<28\% \\
\bottomrule
\end{tabular}}
    \label{tab:utilization}
    \vspace{2mm}
\end{table}

\subsection{Evaluation on Robustness}
There are mainly two cases where the codewords change drastically. The first case is at the very beginning of the compression process. The offline codewords at the beginning may be largely different from the actual codewords. The second case is during the compression process, the statistics of input data chunk suddenly change. Therefore, we need to evaluate the response time of our system to these cases, which is the duration that our system needs to recover to a reasonable compression ratio.

\begin{figure}[ht]
    \centering
    \vspace{-2mm}
    \includegraphics[width=\linewidth]{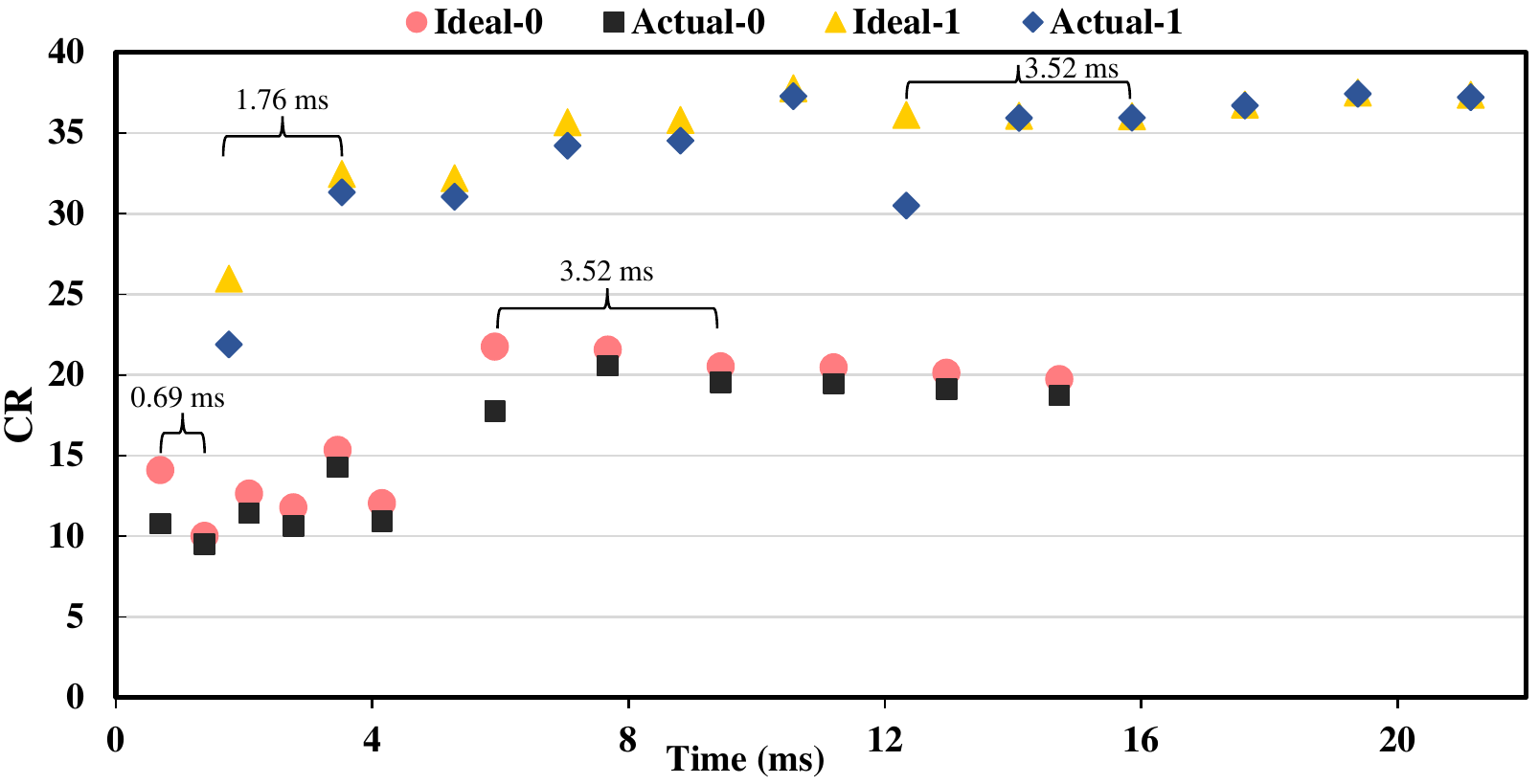}
    \caption{Response time when codewords change drastically.}
    \label{fig:response}
    \vspace{-2mm}
\end{figure}

To prove the robustness of CEAZ, we concatenate two different datasets with different types. Specifically, we concatenate CESM and NYX for the first dataset and concatenate NWChem and S3D for the second dataset. The reason that we concatenate different datasets is that codewords can change drastically both at the beginning and during the processing. Figure \ref{fig:response} shows the response time when current codewords are significantly different from the actual codewords. In the figure, ``Actual-0'' and ``Actual-1'' represent for the actual compression ratio of CEAZ on the two datasets based on offline codewords or previous chunk's codewords, while ``Ideal-0'' and ``Ideal-1'' represent for the ideal compression ratio that CEAZ can achieve on the two datasets using the codewords generated from current chunk. We note that CEAZ can recover to a reasonable compression ratio with only 3.52 $m$s with a strong robustness.

\subsection{Evaluation on Offline Codewords and Codewords Update Frequency}
\label{subsec:evaluation-update}
\begin{figure}[t]
    \centering
%    \vspace{2mm}
    \includegraphics[width=0.95\linewidth]{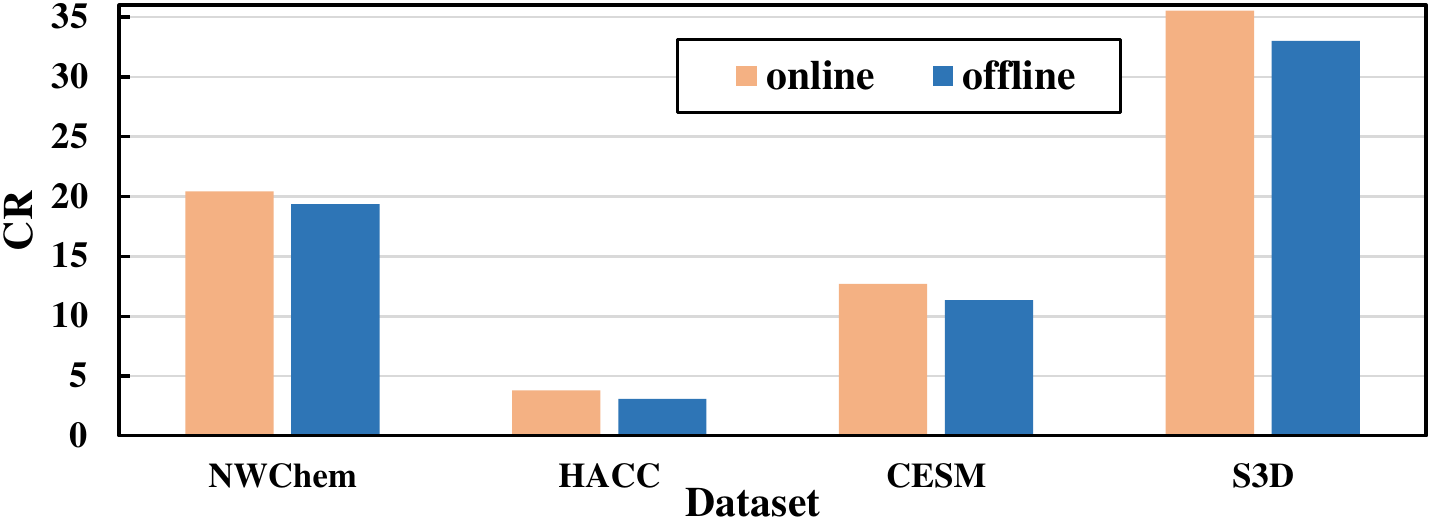}
    \caption{Comparison of compression ratio (CR) between offline codewords and online codewords.}
    \label{fig:offline}
%    \vspace{-1mm}
\end{figure}

We use predefined (offline) codewords at the beginning of the compression process. To prove the effectiveness of our predefined codewords, we evaluate four datasets (i.e., NWChem, HACC, CESM, and S3D) using these codewords with the same value-ranged-based relative error bound of 1e-4 and compare their compression ratios with the optimal ones, as shown in Figure~\ref{fig:offline}.
The orange bars represent the ideal compression ratio achieved by first building Huffman tree and then generating accurate codewords. The blue bars represent the compression ratio by directly using our offline codewords. The compression ratio drops on NWChem, CESM, and S3D are 5.1\%$\sim$10.7\%. The compression ratio degradation on HACC is more obvious (i.e., $\sim$18.2\%); this is because the Lorenzo predictor has low efficiency on HACC dataset, causing many outliers (unpredictable data points), and the distribution of quantization codes generated by the Lorenzo predictor is not statistically representative.

We note that even though HACC's compression ratio drops more than 18.2\%, the compression ratio over $3\times$ can still relieve the communication pressure to a certain extent thanks to our high throughput (will be discussed in Section \ref{subsec:throughput}). 

As aforementioned, frequently updating codewords will decrease the compression ratio due to the overhead of saving codewords. We evaluate the impact of update frequency on the final compression ratio. We perform the experiments on both CESM and NYX. We set the error bound to the value-range-based relative error bound of 1e-4. We choose to update the codewords every 1 MB, 2 MB, 4 MB, 16 MB, 32 MB, 64 MB, 128 MB, 256 MB, and 512 MB. We find that the compression ratio is significantly reduced when the update size is smaller than 32, because the overhead of storing the codewords is relatively large. Moreover, we also observe that the compression ratio decreases when the update size is larger than 256 MB. The reason is that the codewords are outdated to reflect the current symbols frequencies. Therefore, we choose 32 MB as our default update size. 

\subsection{Evaluation on Change of Standard Deviation of Symbol Frequencies}
\label{sec:eval-std}
\begin{figure}[ht]
    \centering
%    \vspace{-1mm}
    \includegraphics[width=0.95\linewidth]{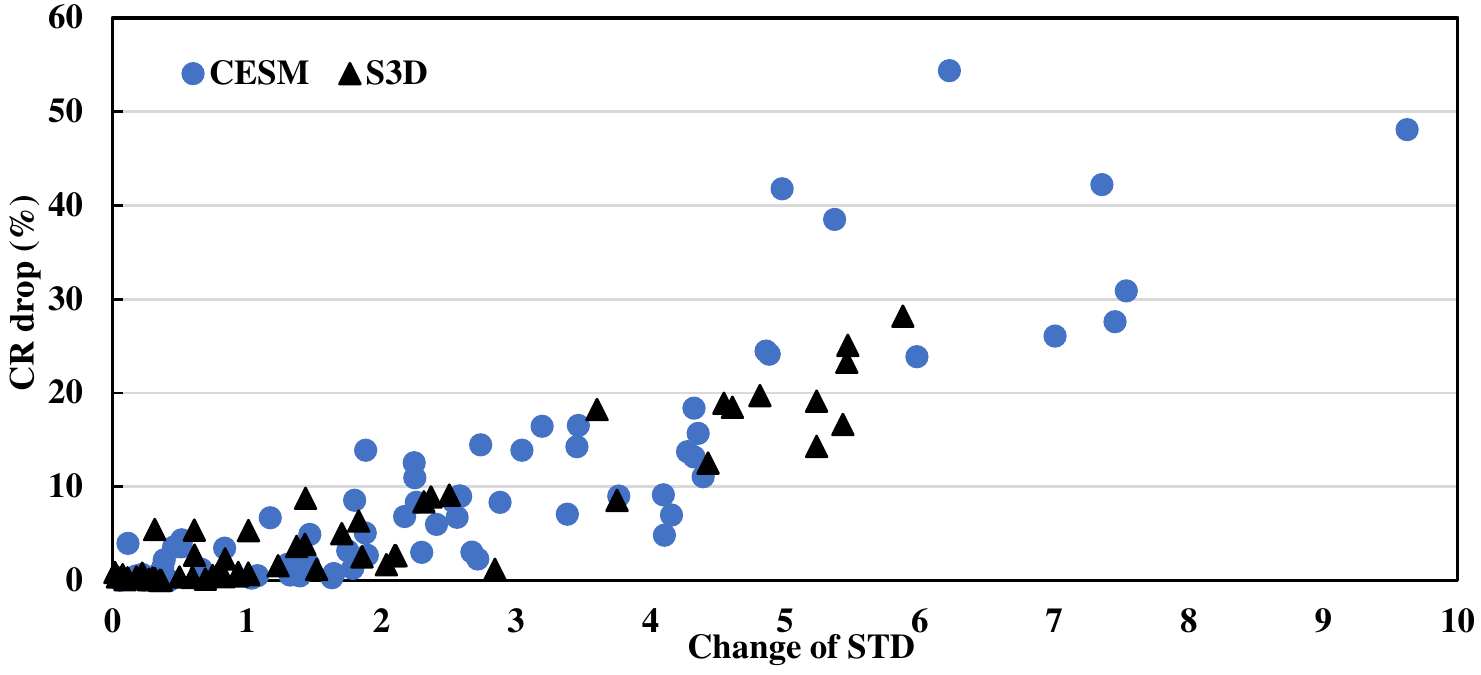}
    \caption{Compression ratio drops with different changes of STD.}
    \label{fig:tau}
%    \vspace{-1mm}
\end{figure}

As aforementioned, we use the change of standard deviation of symbol frequencies (i.e., $\chi = \mid \sigma_0-\sigma_1 \mid$) to determine when to generate new codewords, use old codewords, or use offline codewords. Using previous codewords under a large $\chi$ (a large difference between current and previous distribution) results in a notable drop of compression ratio, while generating new codewords under a small $\chi$ (a small difference between current and previous distribution) leads to a high overhead of Huffman coding. 
Thus, we use experiments to find the suitable thresholds $\tau_0$ and $\tau_1$. As shown in Figure~\ref{fig:tau}, the drop of ratio is less than 10\% when $\chi \leq 3.05$, while the drop is over 25\% when $\chi \geq 4.88$. So, we set $\tau_0$ and $\tau_1$ to 3.05 and 4.88, respectively, to meet both requirements on ratio and performance.

\subsection{Evaluation on Fixed-Ratio Mode}
\label{sec:eval-fix-ratio}
\begin{figure}[ht]
    \centering
    % \vspace{1mm}
    \includegraphics[width=0.9\linewidth]{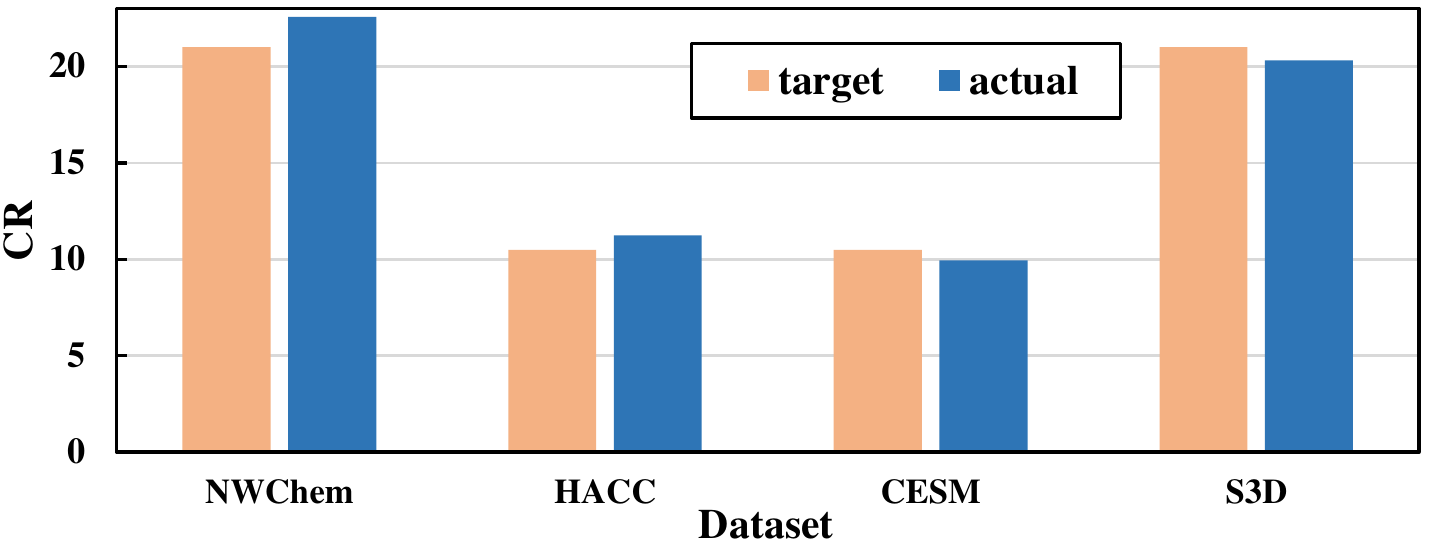}
    \caption{Comparison of target \& actual ratio in fixed-ratio mode.}
    \label{fig:control}
%    \vspace{-1mm}
\end{figure}

As discussed in Section~\ref{sec:design}, our novel compression engine has two working modes: fixed-accuracy mode (i.e. error-bounded mode) and fixed-ratio mode (i.e., fixed bit-rate mode). The fixed-ratio mode can allow the system have a consistent throughput for data transfer.
To verify the effectiveness of our fixed-ratio mode, we set the target compression ratios of 10.5 and 21 for single and double floating-point data, respectively. Figure~\ref{fig:control} shows the compression between the target ratio and the actual ratio. The difference is within 7.5\%, which is acceptable in our use case. 

\subsection{Evaluation on Ratio and Distortion}
Compression ratio is defined as the ratio of original data size to the compressed data size. PSNR is a widely used indicator to assess the distortion of data after lossy compression, which is calculated as $\text{PSNR}= 20\cdot \log_{10}\! \big[(d_{\max}\! - d_{\min} )/\text{RMSE}\big]$.
$N$ is the number of data points and $d_{\max}$ and $d_{\min}$ are the maximal and minimal values, respectively. RMSE is the \textit{root mean squared error}, i.e., $\text{\sffamily sqrt}\left(\frac{1}{N}\textstyle\sum_{i=1}^{N} \left(d_i - d_i^{\bullet}\right)^2\!\right)$, where $d_i$ and $d_i^\bullet$ are the original and decompressed data values, respectively. The larger the PSNR, the lower the data distortion, hence more accurate post analysis.

\begin{figure}[ht]
    \centering
%     \vspace{-1mm}
    \includegraphics[width=\linewidth]{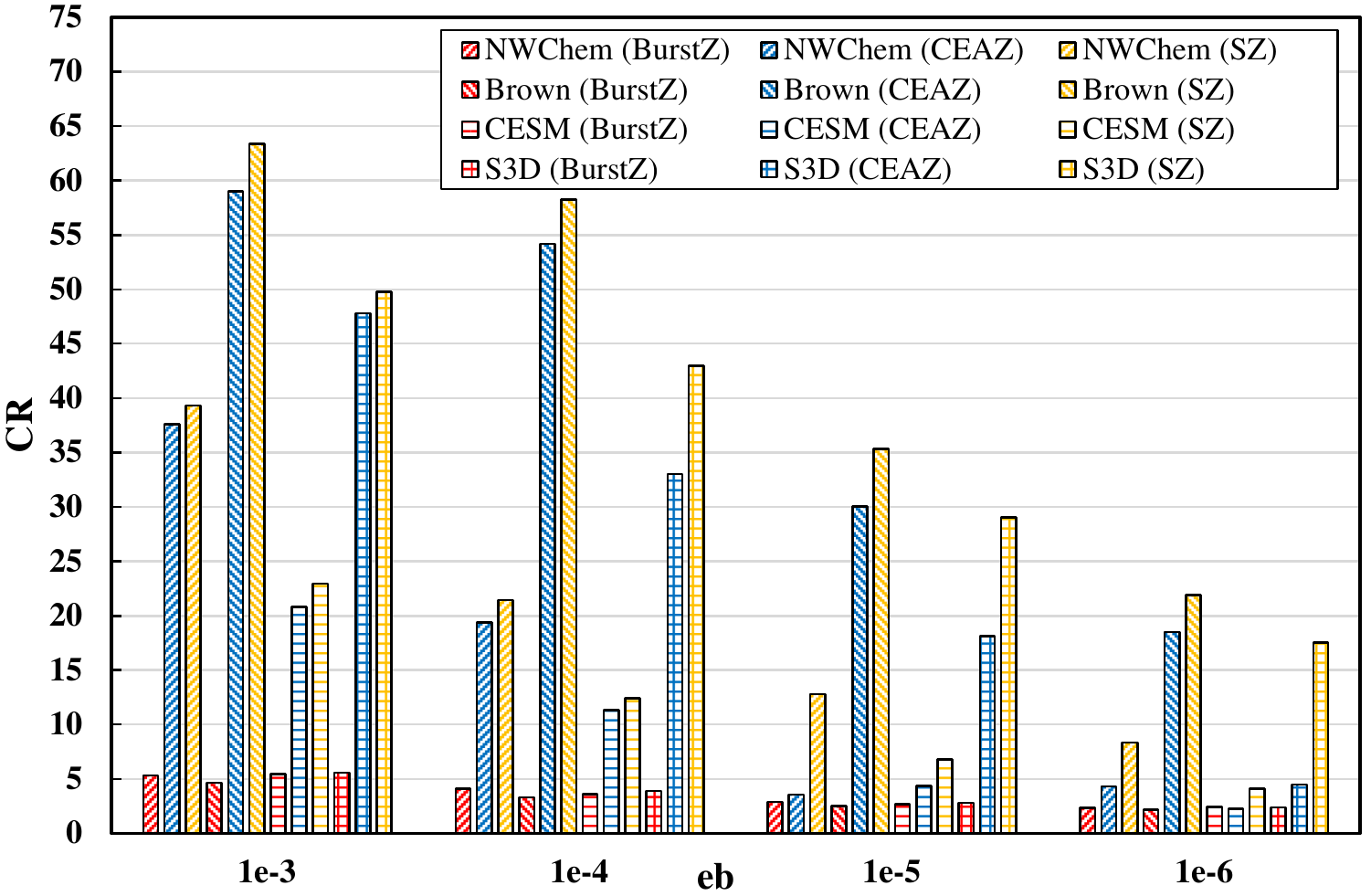}
    \caption{Ratio comparison among BurstZ, CEAZ, and SZ.}
    \label{fig:ratio}
%    \vspace{-1mm}
\end{figure}

Figure~\ref{fig:ratio} shows the comparison of compression ratio among BurstZ, CEAZ, and CPU-SZ on our test datasets with different value-ranged-based relative error bounds of 1e-3$\sim$1e-6. The compression ratio of our CEAZ is notably higher than that of BurstZ. CEAZ consistently provides $1.2\times$$\sim$$16.4\times$ higher compression ratio than BurstZ under the same error bound. Particularly, our CEAZ improves the compression ratio by up to $12\times$ on the Brown dataset over BurstZ when the error bound is equal to 1e-3. Compared to the CPU-SZ, the degradation of compression ratio is within 23.3\% on all test datasets and reasonable error bounds. This result demonstrates the effectiveness of our adaptive strategy and offline codewords.

We also compare the compression ratio and throughput among LZ4, Gzip, and CEAZ with the best compression mode for LZ4/Gzip. Table~\ref{tab:lossless} illustrates that CEAZ is effective to reduce the scientific data size with high throughput.

In addition, Table~\ref{tab:psnr} shows the comparison of distortion (PSNR) between CEAZ and CPU-SZ under different error bounds. The degradation of PSNR is within 4 dB under very high PSNRs (all higher than 60 dB).

\begin{table}[ht]
    \vspace{-1mm}
    \caption{Comparison of compression ratio and averaged throughput (in GB/s) among LZ4, Gzip, CEAZ, and CPU-SZ on test datasets.}
    \centering
    \resizebox{0.9\linewidth}{!}{\begin{tabular}{@{} >{\bfseries}l *{6}{r} @{}}
\toprule
   &eb &NWChem &Brown &CESM &S3D & throughput\\ 
\midrule
LZ4 &N/A &1.005 &1.003 &1.182 &1.055 &1.43 \\
Gzip &N/A &1.056 &1.442 &1.361 &1.181 &1.14 \\ 
CEAZ &1e-4 &20.4 &58.2 &12.3 &35.0 &17.8 \\
CPU-SZ &1e-4 &21.5 &58.2 &12.5 &43.0 &0.31 \\
\bottomrule
\end{tabular}
}
    \vspace{-1mm}
    \label{tab:lossless}
\end{table}

\begin{table}[ht]
    \vspace{-1mm}
    \caption{Distortion (i.e., PSNR) comparison between CEAZ and SZ.}
    \centering
    \resizebox{\linewidth}{!}{\begin{tabular}{@{} >{\bfseries}l *{8}{r} @{}}
\toprule
\bfseries\itshape eb &
\multicolumn{2}{r}{\bfseries NWChem} &
\multicolumn{2}{r}{\bfseries Brown} &
\multicolumn{2}{r}{\bfseries CESM} &
\multicolumn{2}{r}{\bfseries S3D} 
\\
\cmidrule(lr){2-3}
\cmidrule(lr){4-5}
\cmidrule(lr){6-7}
\cmidrule(l){8-9}
   &SZ   &CEAZ  &SZ &CEAZ &SZ &CEAZ &SZ &CEAZ\\
\midrule
1e-3  &65.8  &75.1  &64.7  &64.8  &65.6  &65.4  &71.2  &68 \\
1e-4  &85.8  &90.4  &84.8  &84.8  &85.4  &84.8  &88.8  &84.9 \\
1e-5  &105.6 &107.7 &104.8 &104.8 &105.4 &105.3 &108.2 &104.8 \\
1e-6  &125.0   &126.0   &124.8 &124.8 &125.5 &125.3 &127.7 &124.8 \\
\bottomrule
\end{tabular}
}
     \vspace{-2mm}
    \label{tab:psnr}
\end{table}

\subsection{Evaluation on Time, Throughput, Latency}
\label{subsec:throughput}
\paragraph{Time} The compression time (excluding the file loading and dumping time) is measured as the period from the moment that FPGA receives the data through the moment that the whole compression is finished with output bytes. We show the comparison of compression time among BurstZ, CEAZ, and CPU-SZ in Table~\ref{tab:compression_time}. The error bounds are 1e-4 and 1e-5. We observe that CEAZ reduces the compression time on average 55.8\% compared with the second-best BurstZ on the same dataset. 

\begin{table}[ht]
    \vspace{-1mm}
    \caption{Compression time (in second) of different compressors.}
    \centering
    \resizebox{0.8\linewidth}{!}{\begin{tabular}{@{} >{\bfseries}l *{5}{r} @{}}
\toprule
       &eb  &NWChem &Brown &CESM &S3D\\ 
\midrule
BurstZ &1e-4 &1.65 &0.09 &0.23 &6.40\\
BurstZ &1e-5 &1.70 &0.13 &0.35 &9.67\\
CEAZ   &1e-4 &0.67 &0.04 &0.11 &2.93\\
CEAZ   &1e-5 &0.67 &0.04 &0.11 &2.93\\
CPU-SZ &1e-4 &27.6 &1.58 &12.28 &141.4\\
CPU-SZ &1e-5 &28.6 &1.59 &12.37 &160.8\\
\bottomrule
\end{tabular}}
    \vspace{-1mm}
    \label{tab:compression_time}
\end{table}

\begin{figure}[t]
    \centering
    % \vspace{-1mm}
    \includegraphics[width=\linewidth]{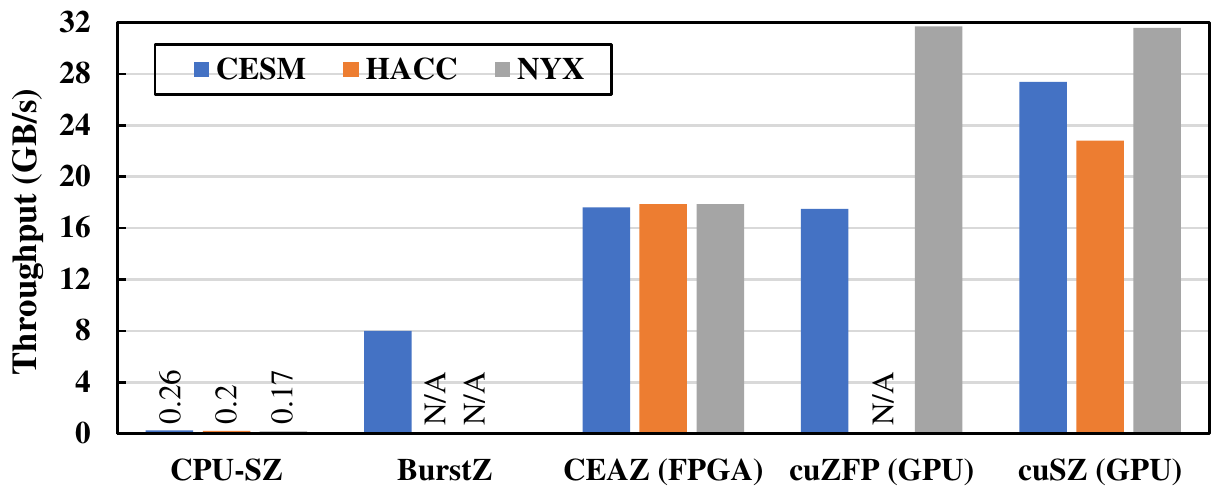}
    \caption{Throughput comparison among BurstZ, CEAZ, CPU-SZ.}
    \label{fig:throughput}
%    \vspace{2mm}
\end{figure}

\paragraph{Throughput} The compression throughput is defined as the size of data being compressed per second. In order to compare the compression throughputs on CPU and GPU, we evaluate the throughputs of CPU-SZ, cuSZ, cuZFP, and CEAZ across three datasets, as shown in Figure~\ref{fig:throughput}. We set the error bound to 1e-4, which has the data distortion (i.e., PSNR) of about 85 dB.

Note that the throughput of cuZFP is highly related to its user-set fixed bitrate according to the previous study \cite{jin2020understanding}, whereas the throughputs of BurstZ, cuSZ, and CEAZ are almost unaffected by the user-set error bound. Therefore, we choose the acceptable fixed bitrate for cuZFP, which generates the data distortion similar to that of CEAZ. The figure illustrates that CEAZ can consistently provide about 17.8 GB/s throughput with different error bounds, which is $2.3\times$ higher than BurstZ on average. Compared with the serial CPU-SZ, CEAZ improves the throughput by $67.4\times$ on average.

It is worth noting that CEAZ can provide at least 3.0$\times$ higher compression ratio compared to BurstZ when error bounds are from 1e-3 to 1e-4. Due to such a high compression ratio (or high data reduction capability), the bandwidth of dumping compressed data (even with the smallest ratio) is still less than the bandwidth capacity of the Ethernet transceiver in our FPGA board (i.e., 100 Gb/s), thereby, the overall throughput has not been bounded by this capacity. Moreover, our clock frequency is around 300 MHz; thus, the throughput could be further improved by increasing the frequency. 

\begin{table}[ht]
    \vspace{-2mm}
    \caption{Hardware specifications of tested GPU and FPGA.}
    \centering\small
    \begin{tabular}{@{} rrrrrr @{}}
\toprule
%  \multicolumn{2}{r}{FPGA}
% & \multicolumn{2}{r}{GPU}
% \\
% \cmidrule(r){1-2}
% \cmidrule(l){3-4}
%       chip &board &chip &card \\ 
% \midrule
% 4w  &30w &65w &300w \\
     &\textbf{Power} &\textbf{Tech.} &\textbf{BW} &\textbf{Freq.} &\textbf{Peak Perf.} \\
\midrule
V100 &250W &12 nm &900 GB/s &1.25 GHz &15.7 TF/s \\
U280 &225W &16 nm &460 GB/s &0.3 GHz &- \\
\bottomrule
\end{tabular}
    \vspace{-2mm}
    \label{tab:hardware}
\end{table}

CEAZ stably provides a compression throughput of 17.8 GB/s, which is about 56\%$\sim$78\% of cuSZ/cuZFP's throughputs (note that cuZFP has a fairly low compression quality on 1D HACC). CEAZ and cuSZ/cuZFP are implemented on Xilinx Alveo U280 FPGA card and Nvidia Tesla V100 GPU, respectively.
Table~\ref{tab:hardware} lists the hardware specifications of V100 GPU and U280 FPGA. Note that V100 provides up to 900 GB/s bandwidth (1.96$\times$ higher than U280) and 1.25 GHz frequency (4.2$\times$ higher than U280). In addition, although we cannot find U280's theoretical peak performance, a similar FPGA, Intel S10 NX, has the peak performance of 3.96 TF/s (4.0$\times$ lower than V100) \cite{boutros2020beyond,nguyen2020performance}. Therefore, FPGA-based CEAZ is more efficient in resource utilization than GPU-based cuSZ/cuZFP.

We also note that there is a recent work (called DE-ZFP) \cite{habboush2022zfp} that develops an FPGA implementation of a modified ZFP algorithm. However, the paper only evaluates DE-ZFP on three datasets with two absolute error bounds. Thus, we select their tested data fields that have reasonable relative error bounds (under the absolute error bounds of 1e-3 and 1e-6) and perform a comparison. Our evaluation shows that CEAZ has 1.7$\times$ higher compression ratio and 11.1$\times$ higher compression throughput over DE-ZFP on average. 

\begin{table}[t]
%    \vspace{-2mm}
    \caption{Latency ($\mu$s) of different lossy compressors on small data.}
    \centering
    \resizebox{0.95\linewidth}{!}{\begin{tabular}{@{} >{\bfseries}l *{4}{r} @{}}
\toprule
       &CPU-SZ &cuSZ (GPU) &cuZFP (GPU) &CEZA (FPGA) \\ 
\midrule
1 KB &69 &358.3 &16.6 &3.7 \\
4 KB &114 &416.6 &19.4 &3.8 \\
16 KB &147 &507.4 &27.4 &4.5 \\
64 KB &458 &546.6 &46.7 &7.0 \\
\hline
\end{tabular}
}
    \vspace{2mm}
    \label{tab:latency}
\end{table}

\paragraph{Latency} We evaluate the latency of CPU-SZ, cuSZ, cuZFP, and CEAZ on small datasets to demonstrate the capability of using CEAZ in reducing the communication cost in future work, as shown in Table~\ref{tab:latency}.
The test small data are chunked from the CESM-ATM dataset. The table illustrates that CEAZ achieves up to 113.4$\times$ and 6.7$\times$ lower latency than cuSZ and cuZFP, respectively.

\subsection{Parallel Performance Evaluation}
We demonstrate the parallel performance in two ways. We first evaluate the throughput of CEAZ with multiple pipelines in the single FPGA board. We then evaluate CEAZ with multiple nodes.

\subsubsection{Multi-pipeline Evaluation}
\begin{figure}[ht]
    \centering
    \vspace{-2mm}
    \includegraphics[width=\linewidth]{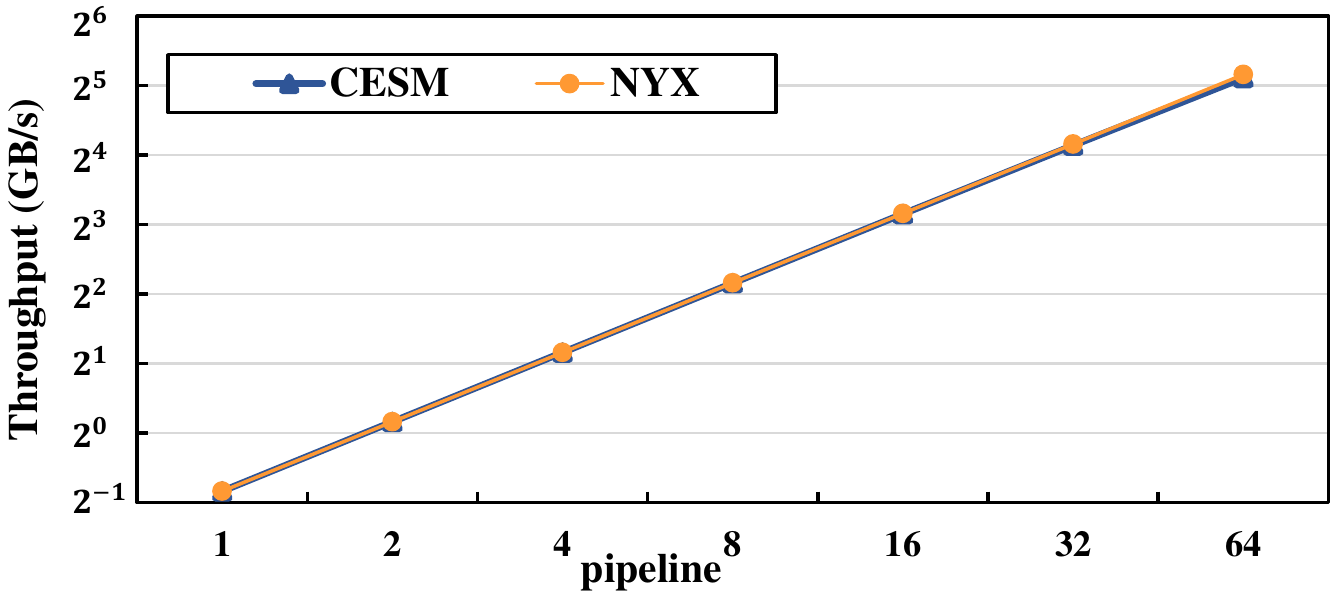}
    \caption{Compression throughputs with multiple pipelines.}
    \label{fig:multi}
    \vspace{-4mm}
\end{figure}

We choose the CESM-ATM and NYX datasets and set the value-range-based error bound to 1e-4, which is commonly used in the CESM and NYX applications \cite{poppick2020statistical}. We increase the compression pipelines from 1 to 64. Figure~\ref{fig:multi} illustrates that the throughput (in log-scale) increases linearly as the number of pipelines increases.
CEAZ can achieve this high scalability because \Circled{1} our compression engine reads data from HMB2 with a very high bandwidth of 460 GB/s, and \Circled{2} our compression engine adopts dual-quant to fully remove the data dependency so that we can process different chunks of the dataset in parallel. 

\begin{figure*}[]
    \includegraphics[width=\linewidth]{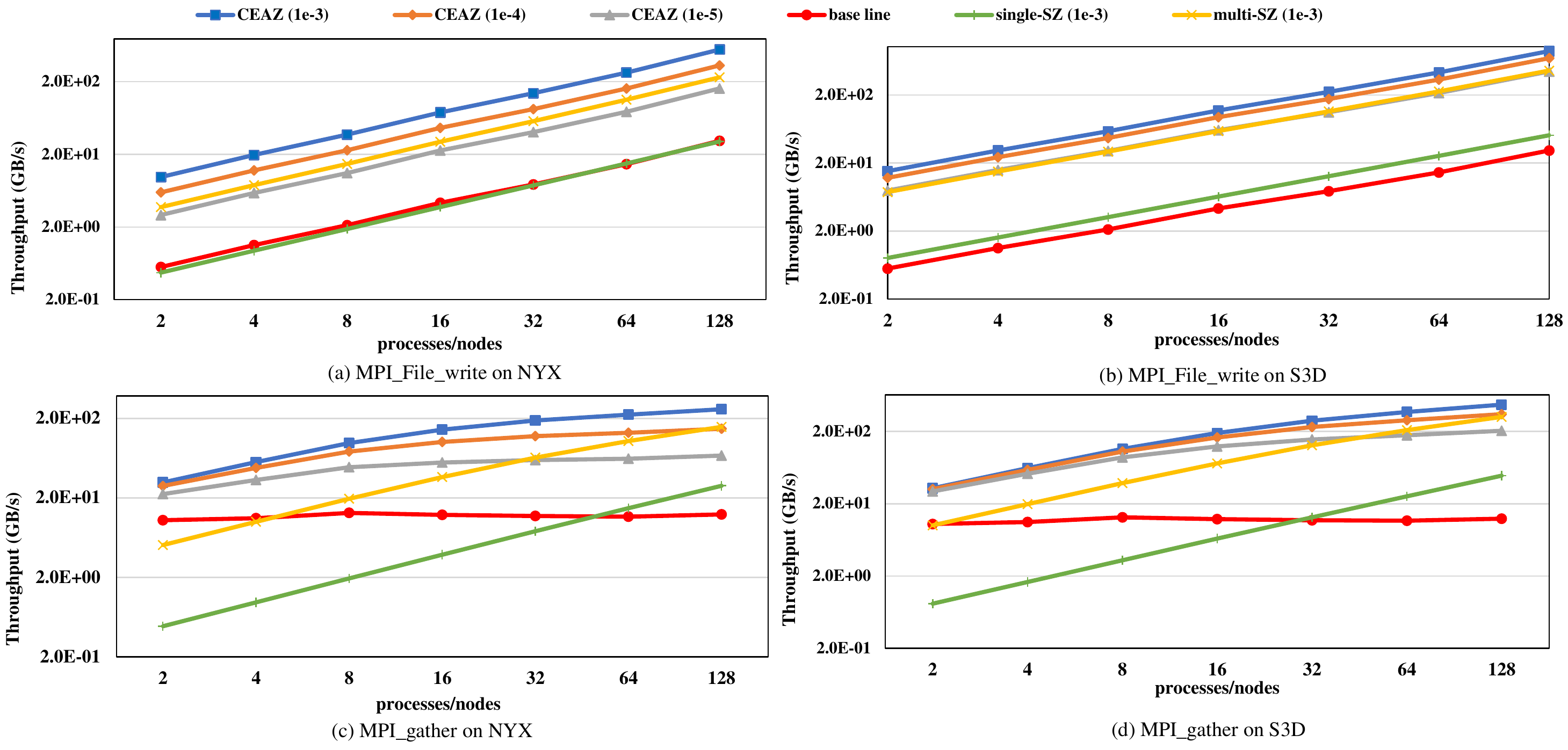}
%    \vspace{-2mm}
    \caption{Throughput comparison of CEAZ-accelerated and original MPI\_File\_write and MPI\_Gather on NYX and S3D with different numbers of processes/nodes. ''single-SZ'' denotes SZ with a single CPU core/node. ''multi-SZ'' denotes SZ with 32 CPU cores/node.}
    \label{fig:mpi-ceaz}
%    \vspace{-2mm}
\end{figure*}

\subsubsection{Multi-node Evaluation}
We evaluate the performance improvements of MPI-IO and MPI collective operations gained from CEAZ, i.e., MPI\_File\_write and MPI\_Gather. We conduct our experiments with up to 128 nodes (one process per node). Each node holds a copy of datasets for compression and transmission, i.e., 3.0 GB of NYX and 10.2 GB of S3D per node. Thus, the overall data size for parallel I/O is up to 1.3 TB with 128 nodes.
We evaluate CPU-SZ on both a single core and 32 cores with the error bound of 1e-3. Table~\ref{tab:compression_ratio0} shows the compression ratio of CPU-SZ and CEAZ on NYX and S3D with different error bounds.

\begin{table}[ht]
%    \vspace{-1mm}
    \caption{Compression ratio of CPU-SZ and CEAZ on NYX and S3D.}
    \centering
    \resizebox{0.72\linewidth}{!}{\begin{tabular}{ccccc}
\hline
\multicolumn{1}{l}{\multirow{2}{*}{}} & \multicolumn{2}{c}{\textbf{NYX}} & \multicolumn{2}{c}{\textbf{S3D}} \\ \cline{2-5} 
\multicolumn{1}{l}{}                  & CPU-SZ           & CEAZ          & CPU-SZ           & CEAZ          \\ \hline
\textbf{1e-3}                         & 26               & 23.7          & 49.8             & 47.8          \\
\textbf{1e-4}                         & 14.7             & 12.8          & 43.0             & 33            \\
\textbf{1e-5}                         & 7.7              & 5.7           & 29.1             & 18.1          \\ \hline
\end{tabular}}
    \vspace{1mm}
    \label{tab:compression_ratio0}
\end{table}

Figure~\ref{fig:mpi-ceaz} (a) and Figure~\ref{fig:mpi-ceaz} (b) show the MPI-IO throughputs on the NYX and S3D datasets with different approaches. 
The throughput of original MPI\_File\_write (without compression) increases as the number of nodes increases and can reach up to 30.5 GB/s with 128 nodes in Summit, as shown as the baselines in the figures.
The single-core-SZ-supported MPI\_File\_write only achieves an overall throughput of 51.0 GB/s (including compression time and time to write compressed data) on S3D when using 128 nodes, which is 67.3\% higher than the baseline.
This is because the compression throughput of single-core CPU-SZ (i.e., about 0.41 GB/s) is not fast enough compared to the state-of-the-art interconnect such as InfiniBand HDR with a bandwidth of 200Gb/s.
Note that unlike on S3D, the single-core-SZ-supported MPI\_File\_write on NYX is slower than the baseline, since the compression ratio of SZ on NYX (i.e., 26.0) is much lower than S3D (i.e., 49.8). 
The multi-core-SZ-supported MPI\_File\_write can provide an overall throughput of up to 456.56 GB/s when using 128 nodes, which is 15.0$\times$ higher than the baseline.
In comparison, CEAZ-supported MPI\_File\_write can improve the overall throughput (including compression time and time to write compressed data) by 18.0$\times$ and 28.9$\times$ on NYX and S3D, respectively. 

Figure~\ref{fig:mpi-ceaz} (c) and Figure~\ref{fig:mpi-ceaz} (d) show the MPI\_Gather throughputs on the NYX and S3D datasets with different approaches.
The throughput of original MPI\_Gather reaches 12.4 GB/s with 128 nodes in Summit, as shown as the baselines in the figures. 
Similar to MPI\_File\_write, the single-core-SZ-supported MPI\_Gather only achieves an overall throughput of up to 48.7 GB/s with 128 nodes, which is just 3.9$\times$ higher than the baseline; the multi-core-SZ-supported MPI\_Gather can provide an overall throughput of up to 316.9 GB/s when using 128 nodes, which is 25.6$\times$ higher than the baseline.
In comparison, CEAZ-supported MPI\_Gather can improve the overall throughput by $21.0\times$ and $37.8\times$ on NYX and S3D, respectively, due to the high efficiency of CEAZ.  

\section{Conclusion and Future Work}
\label{sec:conclusion}

In this work, we propose CEAZ: a hardware-algorithm co-design of efficient and adaptive lossy compressor for scientific data. To achieve both high compression ratio and throughput, we propose an efficient Huffman coding approach that can adaptively update Huffman codewords online based on our offline generated representative Huffman codewords. We also derive a theoretical analysis to accurately control compression ratio under the error-bounded compression mode, enabling an accurate generation of offline Huffman codewords and a fixed-ratio compression mode. Our evaluation demonstrates that CEAZ outperforms the second-best FPGA-based error-bounded lossy compressor by 2.3$\times$ of throughput and 3.0$\times$ of compression ratio. CEAZ improves MPI\_File\_write and MPI\_Gather by up to 28.9$\times$ and 37.8$\times$, respectively, with 128 nodes in Summit. 

In future work, we plan to deploy our system to FPGA-based clusters and extend CEAZ to DPU-based systems.
\section*{Acknowledgment}
%\small 
This work was partially supported by the National Science Foundation OAC-2042084 and OAC-2104024. This work was also partially supported by the Compute-Flow-Architecture (CFA) project under PNNL’s Data-Model-Convergence (DMC) LDRD Initiative. This research used resources of the Oak Ridge Leadership Computing Facility at the Oak Ridge National Laboratory, which is supported by the Office of Science of the U.S. Department of Energy under Contract No. DE-AC05-00OR22725.
%This work was partially supported by by the U.S. Department of Energy, Office of Science, Office of Advanced Scientific Computing Research, ComPort: Rigorous Testing Methods to Safeguard Software Porting, under Award Number 78284.

\newpage
\bibliographystyle{ACM-Reference-Format}
\bibliography{refs}

\end{document}